\documentclass[pdflatex,sn-mathphys-num]{sn-jnl}

\usepackage{graphicx}%
\usepackage{multirow}%
\usepackage{amsmath,amssymb,amsfonts}%
\usepackage{amsthm}%
\usepackage{mathrsfs}%
\usepackage[title]{appendix}%
\usepackage{xcolor}%
\usepackage{textcomp}%
\usepackage{manyfoot}%
\usepackage{booktabs}%
\usepackage{algorithm}%
\usepackage{algorithmicx}%
\usepackage{algpseudocode}%
\usepackage{listings}%
\usepackage{lineno}

\theoremstyle{thmstyleone}%
\theoremstyle{thmstyletwo}%

\theoremstyle{thmstylethree}%

\begin{document}

\title[Article Title]{A Scale-adaptive Vision Model Links \textit{C. elegans} Neuronal Morphology to Behavior for Neurotoxicity Assessment\vspace{1em}}


\author[1,2]{\sur{Haochao} \fnm{Ying} }\email{haochaoying@zju.edu.cn}\equalcont{These authors contributed equally to this work}
\author[1]{\sur{Shenchong} \fnm{Lv} }\email{22318280@zju.edu.cn}\equalcont{These authors contributed equally to this work}
\author[3,8]{\sur{Yutao} \fnm{Sun}}\email{yutaosun@zju.edu.cn}\equalcont{These authors contributed equally to this work.}
\author[4]{\sur{Zijian} \fnm{Tu}}\email{zijiantu@zju.edu.cn}
\author[1]{\sur{Xufeng} \fnm{Jin}}\email{22418013@zju.edu.cn}
\author[2,4,8]{\sur{Yuyang} \fnm{Xu}}\email{xuyuyang@zju.edu.cn}
\author[1]{\sur{Yizhe} \fnm{Wang}}\email{zdyfwyz597@163.com}
\author[5]{\sur{Wei} \fnm{Yang}}\email{yangwei@zju.edu.cn}
\author*[6]{\sur{Xiaomin} \fnm{Yue}}\email{yuexiaomin@zju.edu.cn}
\author*[2,4,7,8]{\sur{Jian} \fnm{Wu}}\email{wujian2000@zju.edu.cn}
\author*[1]{\sur{Peilin} \fnm{Yu}}\email{yupeilin@zju.edu.cn}

\affil[1]{\orgdiv{School of Public Health and Second Affiliated Hospital}, \orgname{Zhejiang University School of Medicine}, \orgaddress{\city{Hangzhou}, \postcode{310058}, \country{China}}}

\affil[2]{\orgdiv{State Key Laboratory of Transvascular Implantation Devices and TIDRI}, \orgname{Second Affiliated Hospital of Zhejiang University School of Medicine}, \orgaddress{\city{Hangzhou}, \postcode{310009}, \country{China}}}

\affil[3]{\orgdiv{Polytechnic Institute}, \orgname{Zhejiang University}, \orgaddress{\city{Hangzhou}, \postcode{310058}, \country{China}}}

\affil[4]{\orgdiv{College of Computer Science and Technology}, \orgname{Zhejiang University}, \orgaddress{\city{Hangzhou}, \postcode{310027}, \country{China}}}

\affil[5]{\orgdiv{Medical College}, \orgname{Guizhou University}, \orgaddress{\city{Guiyang}, \postcode{550025}, \country{China}}}

\affil[6]{\orgdiv{Department of Biophysics and Department of Neurosurgery}, \orgname{Fourth Affiliated Hospital of Zhejiang University School of Medicine}, \orgaddress{\city{Yiwu}, \postcode{322000}, \country{China}}}

\affil[7]{\orgdiv{Liangzhu Laboratory}, \orgname{Zhejiang University School of Medicine}, \orgaddress{\city{Hangzhou}, \postcode{311113}, \country{China}}}

\affil[8]{\orgdiv{Zhejiang Key Laboratory of Medical Imaging Artificial Intelligence}, \orgaddress{\city{Hangzhou}, \postcode{310058},  \country{China}}}

\abstract{
Neurological disorders are a leading cause of global disability and are increasingly linked to environmental chemical exposures. Yet neurotoxicity assessment still relies on hand-scored morphological readouts that are subjective and poorly predictive of behavioral outcomes.
\textit{Caenorhabditis elegans} provides a genetically tractable, 3R-compliant alternative, but quantifying neuronal phenotypes from confocal microscopy at scale remains computationally challenging: existing vision foundation models, trained on natural or radiological images, cannot resolve the sparse signals and multi-scale lesions of neuronal imaging.
Here, we introduce a dedicated self-supervised vision model for \textit{C. elegans} dopaminergic neurons, together with CeNeuMorph, a multi-grained confocal benchmark of 27,117 annotated images. Specifically, moving beyond standard Masked Autoencoders, we propose a scale-adaptive masked image modeling strategy that jointly learns representations across resolutions and patch sizes under a fixed token budget. By decoupling structural semantic learning from rigid grid constraints, the model effectively resolves the full spectrum of neurodegenerative lesions---ranging from fine dendritic beading to gross soma shrinkage---within a tractable computational framework.
Finally, our model surpasses both generalist (DINOv2, SigLIP2) and biomedical (MedSAM, BiomedCLIP, MedImageInsight) foundation models across classification, segmentation and detection tasks---reaching 88.4\% accuracy on neuronal breakage detection.
Fusing visual features with morphological descriptors enables prediction of dopamine-dependent behavioral deficits (R$^2$=0.498).
Screening 180 agrochemicals, we identify the benzimidazole moiety as a previously unrecognized determinant of dopaminergic neurotoxicity.
Together, the work demonstrates how scale-adaptive self-supervised learning can connect morphology to function for a scalable alternative to mammalian~\textit{in vivo} models for neurotoxicity assessment and drug discovery.
}




\maketitle

\section{Introduction}\label{sec1}



Neurological disorders are the leading cause of global disability and the second leading cause of death, with growing evidence linking many of these conditions to environmental chemical exposures~\cite{bib52,bib53}. The environmental dimension underscores the need for a reliable assessment of neurotoxicity to identify functionally relevant neuronal damage. However, quantitative assessment of neuronal integrity remains limited, as current approaches provide only hand-scored or subjective morphological readouts that rarely capture the structural alterations underlying behavioral deficits~\cite{bib54}. This gap calls for scalable and objective approaches that systematically characterize neuronal morphology and link it to behavioral outcomes, thereby supporting neurotoxicity assessment, drug discovery, and environmental risk evaluation~\cite{bib55}.

\textit{Caenorhabditis elegans} offers a uniquely well-suited platform to meet this need: 60-80\% of its genes have human orthologues, its nervous system shares fundamental architectural principles and signaling pathways with mammals, and its transparency, rapid reproduction, well-defined lineage, and low maintenance cost align naturally with the 3R (Replacement, Reduction, Refinement) principles~\cite{bib56,bib57,bib58}.
Functional studies in this organism typically combine morphological imaging, which reports structural changes in neurons, with behavioral assays that reflect the functional output of neural circuits~\cite{bib59}.
Both readouts, however, are limited by entrenched methodological weaknesses: morphological scoring remains largely manual and subjective, and behavioral assays lack standardized statistical pipelines~\cite{hart2006behavior}, leaving the structure–function relationship largely uncharacterized at scale.
Computer vision offers a principled, labor-efficient, and reproducible alternative to manual scoring, but existing bioimage models have largely been developed for bright-field or video data and rarely exploit the optical sectioning and subcellular resolution of confocal fluorescence microscopy---the modality of choice for resolving fine neuronal morphology in \textit{C. elegans}.

Although vision foundation models (FMs) have reshaped medical image analysis~\cite{he2024foundation, ma2024segment, zhang2024biomedclip}, transferring them to confocal neurotoxicity imaging confronts two coupled obstacles.
First, a severe domain shift exists: 2D maximum intensity projections from 3D confocal microscopy exhibit sparse fluorescent signals that diverge fundamentally from standard FM training distributions~\cite{weigert2018content, dutta2024recent}. This gap precludes the direct transfer of representations required to resolve fine-grained degenerative changes like dendritic punctate morphology (e.g., beading). Second, while Masked Image Modeling (MIM) mitigates label scarcity~\cite{he2022masked, xie2021simmim}, standard fixed-size masking fails to capture multi-scale neuronal pathologies ranging from localized anomalies within a single patch to widespread damage spanning the entire image. This imposes a rigid spatial trade-off: small patches trigger computational prohibitive during high-resolution downstream fine-tuning, whereas large patches inevitably obscure critical fine-grained details~\cite{beyer2023flexivit, li2022exploring}. Consequently, scalable neurotoxicity screening demands a scale-adaptive framework that decouples feature learning from fixed grids to capture both local lesions and global deformations within a tractable computational budget.

Here we present a scale-adaptive self-supervised learning framework for dopaminergic neurotoxicity screening in \textit{C. elegans}, together with CeNeuMorph, a purpose-built benchmark of 27,117 multi-task annotated confocal images. 
Departing from standard Masked Autoencoders, we introduce a masked image modeling strategy that jointly learns across resolutions and patch sizes under a fixed token budget, allowing the model to capture morphological alterations spanning the full damage spectrum---from few-pixel dendritic beading to gross soma shrinkage. After scale-adaptive pre-training, the encoder is fine-tuned with a dual-scale feature pyramid for downstream classification, segmentation, and detection tasks. We further fuse the encoder with semantic morphological descriptors to predict dopamine-dependent behavioral deficits---closing the structure–function loop that has historically been treated as two disconnected readouts. Applying this framework to screen 180 agrochemicals, we identify the benzimidazole moiety---a previously unrecognized structural determinant of dopaminergic neurotoxicity. Together, our work delivers both a methodological advance---a confocal-native, scale-aware vision model with an open benchmark---and a mechanistically interpretable platform that offers a scalable alternative to mammalian~\textit{in vivo} models for neurotoxicity assessment.

\begin{figure}[!htbp]
\centering
\includegraphics[width=1.0\textwidth, trim=0cm 3.4cm 0cm 0cm, clip]{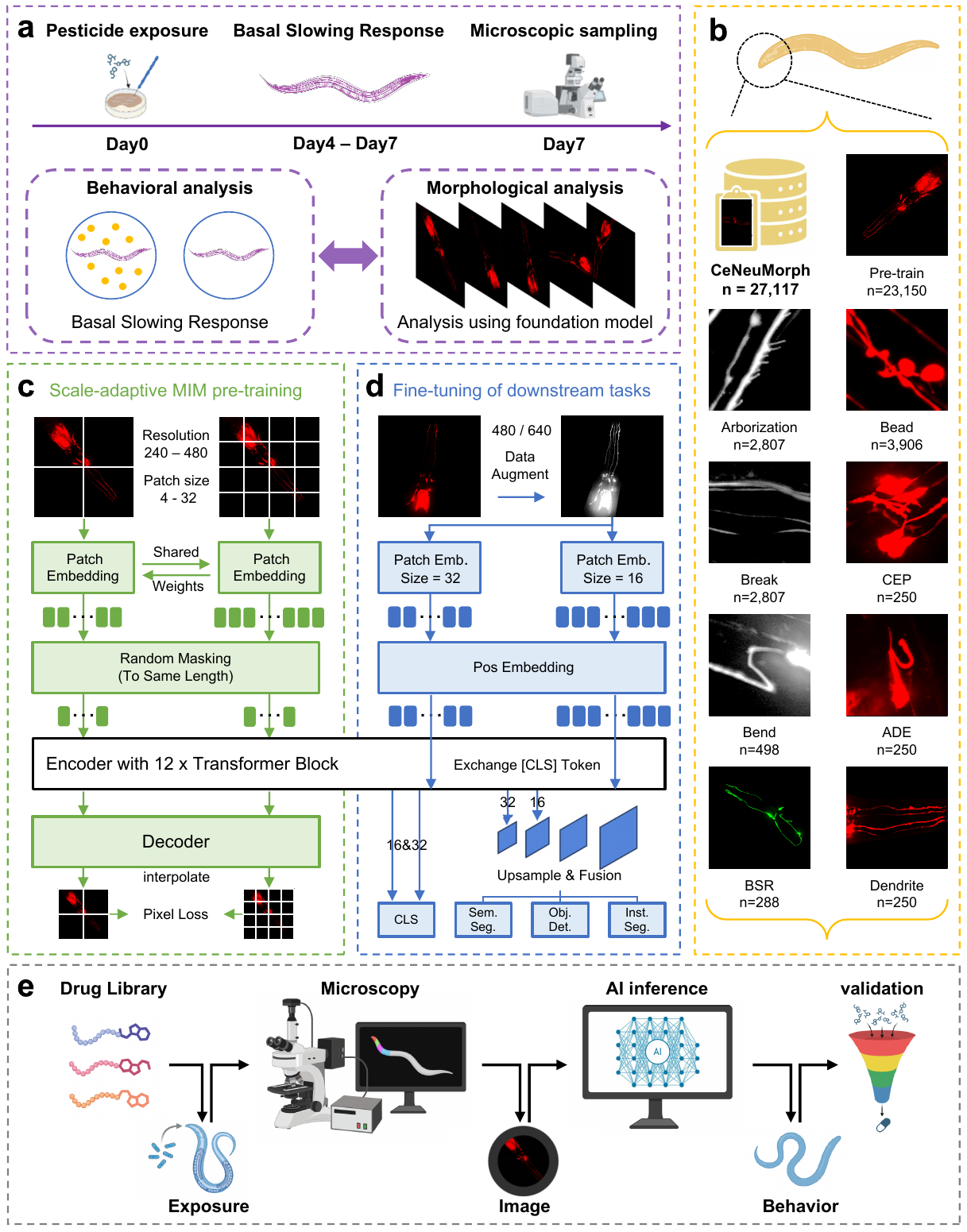}
\caption{\textbf{Overview of the study.} \textbf{a}, Experimental pipeline integrating pesticide exposure (Day 0) with behavioral (BSR) and morphological sampling (Day 7). \textbf{b}, The CeNeuMorph dataset ($n=27,117$), providing multi-task benchmarks. \textbf{c}, Scale-adaptive Masked Image Modeling (MIM) pre-training framework. \textbf{d}, Dual-scale fusion strategy for downstream adaptation, bridging local textural details with global geometry. \textbf{e}, Integrated drug-screening workflow, from library exposure to AI-driven validation and behavioral correlation.}\label{fig1}
\end{figure}

\section{Results}\label{sec2}

\begin{sidewaystable}[htbp]
    \centering
    \caption{Performance benchmarking of the scale-adaptive framework against state-of-the-art pre-training methods and vision foundation models. We evaluate our model across multi-grained phenotypic downstream tasks, including image-level classification of morphological anomalies, semantic segmentation of dendritic beading, and instance-level detection and segmentation of neuronal structures. Compared to established self-supervised pre-training baselines and large-scale generalist/biomedical foundation models, our model demonstrates superior performance across the majority of metrics. This advantage is particularly evident in resolving subtle pathological phenotypes, such as neuronal breakage and precise somata boundaries. Performance is evaluated using Accuracy for classification, Dice score for semantic segmentation, bounding box mAP (mAP$^{\text{box}}$) for object detection, and mask mAP (mAP$^{\text{mask}}$) for instance segmentation.}
    \label{tab:model_comparison}
    \footnotesize
    \setlength{\tabcolsep}{4pt}
    \begin{tabular}{llcccccccc}
    \toprule
    \multirow{2}{*}{Backbone} & \multirow{2}{*}{Method or Model} & \multirow{2}{*}{Train Data / Res.} & \multicolumn{3}{c}{Classification} & Sem. Seg. & Obj. Det. & \multicolumn{2}{c}{Inst. Seg.} \\
    \cmidrule(lr){4-6} \cmidrule(lr){7-7} \cmidrule(lr){8-8} \cmidrule(lr){9-10}
     & & & Break & Arb. & Bend & Bead & Dendrite & ADE & CEP \\
    \midrule
    \multicolumn{10}{l}{\textit{Pre-training Methods}} \\
    ViT-B & MoCo v3 (ICCV '21) & CeNeuMorph / 224 & 86.40$_{\pm0.78}$ & 81.80$_{\pm0.51}$ & 76.84$_{\pm3.17}$ & 76.34$_{\pm0.58}$ & 85.38$_{\pm0.97}$ & 54.00$_{\pm0.92}$ & 63.70$_{\pm1.00}$ \\
    ViT-B & MAE (CVPR '22) & CeNeuMorph / 224 & 86.97$_{\pm1.03}$ & 81.04$_{\pm1.46}$ & 75.26$_{\pm3.00}$ & 75.60$_{\pm0.41}$ & 87.40$_{\pm0.86}$ & 59.32$_{\pm0.40}$ & 66.40$_{\pm1.02}$ \\
    ViT-B & MSN (ECCV '22) & CeNeuMorph / 224 & 83.13$_{\pm0.84}$ & 78.82$_{\pm0.59}$ & 70.26$_{\pm3.67}$ & 73.71$_{\pm0.47}$ & 83.70$_{\pm1.26}$ & 54.00$_{\pm1.09}$ & 61.22$_{\pm0.98}$ \\
    Swin-B & SimMIM (CVPR '22) & CeNeuMorph / 224 & 78.27$_{\pm0.67}$ & 72.26$_{\pm1.06}$ & 76.56$_{\pm0.00}$ & 75.25$_{\pm0.46}$ & 89.40$_{\pm0.96}$ & 58.84$_{\pm0.55}$ & 66.40$_{\pm1.03}$ \\
    Swin-B & MixMAE (CVPR '23) & CeNeuMorph / 224 & 85.12$_{\pm0.31}$ & 81.28$_{\pm0.89}$ & 79.47$_{\pm2.39}$ & 74.18$_{\pm0.48}$ & \textbf{91.56}$_{\pm1.10}$ & 59.78$_{\pm1.38}$ & 68.22$_{\pm0.55}$ \\
    HiViT-B & HiViT (ICLR '23) & CeNeuMorph / 224 & 85.26$_{\pm0.61}$ & 79.19$_{\pm0.11}$ & 78.95$_{\pm3.09}$ & 75.27$_{\pm0.33}$ & 88.90$_{\pm1.78}$ & 54.94$_{\pm1.30}$ & 63.26$_{\pm1.20}$ \\
    \midrule
    \multicolumn{10}{l}{\textit{Foundation Models}} \\
    ViT-B & EVA-02 (IVC '24) & IN-22K / - & 85.07$_{\pm1.30}$ & 81.19$_{\pm1.07}$ & 75.26$_{\pm2.16}$ & 68.44$_{\pm1.19}$ & 82.22$_{\pm0.66}$ & 40.22$_{\pm2.38}$ & 50.86$_{\pm1.80}$ \\
    ViT-B & DINO v2 (TMLR '24) & LVD-142M / - & 87.25$_{\pm1.05}$ & \underline{83.60}$_{\pm0.95}$ & 79.47$_{\pm1.77}$ & 76.94$_{\pm2.05}$ & 86.70$_{\pm1.02}$ & 48.14$_{\pm1.58}$ & 57.58$_{\pm1.64}$ \\
    ViT-B & SigLIP 2 (arXiv '25) & WebLI-10B / - & 85.26$_{\pm0.42}$ & 80.43$_{\pm1.67}$ & 76.32$_{\pm2.08}$ & 64.09$_{\pm1.18}$ & 79.98$_{\pm0.74}$ & 46.36$_{\pm1.84}$ & 56.34$_{\pm1.34}$ \\
    ViT-B & MedSAM (Nat.Com. '24) & MedSAM-1.5M / - & 75.45$_{\pm1.18}$ & 70.43$_{\pm0.51}$ & 72.37$_{\pm3.72}$ & 68.70$_{\pm0.81}$ & 85.64$_{\pm1.78}$ & 52.42$_{\pm0.70}$ & 64.72$_{\pm1.63}$ \\
    ViT-B & BiomedCLIP (NEJM '25) & PMC-15M / - & 85.02$_{\pm1.15}$ & 81.00$_{\pm0.92}$ & 74.74$_{\pm2.16}$ & 73.07$_{\pm0.46}$ & 86.74$_{\pm0.48}$ & 56.50$_{\pm0.90}$ & 64.34$_{\pm0.72}$ \\
    DaviT & MedImgIns. (arXiv '24) & Multi-Med 3.8M / - & 83.89$_{\pm5.15}$ & 81.14$_{\pm3.73}$ & 74.74$_{\pm1.72}$ & 75.52$_{\pm0.59}$ & 89.44$_{\pm0.57}$ & 57.12$_{\pm1.58}$ & 67.58$_{\pm1.28}$ \\
    \midrule
    \multicolumn{10}{l}{\textit{Our Models}} \\
    ViT-B & Ours & CeNeuMorph / 224 & \underline{87.96}$_{\pm0.46}$ & \textbf{83.79}$_{\pm0.76}$ & \underline{81.05}$_{\pm2.73}$ & \underline{78.22}$_{\pm0.38}$ & \underline{89.84}$_{\pm0.33}$ & \underline{62.70}$_{\pm1.69}$ & \underline{69.70}$_{\pm0.78}$ \\
    ViT-B & Ours & CeNeuMorph / 480 & \textbf{88.44}$_{\pm0.35}$ & 83.36$_{\pm0.86}$ & \textbf{81.58}$_{\pm3.36}$ & \textbf{78.26}$_{\pm0.54}$ & 89.66$_{\pm1.15}$ & \textbf{63.14}$_{\pm1.61}$ & \textbf{70.10}$_{\pm0.86}$ \\
    \bottomrule
    \end{tabular}
\end{sidewaystable}

\subsection{A scale-adaptive architecture achieves state-of-the-art multi-grained morphological phenotyping}

To address the scarcity of annotated micro-images and capture the diverse morphological landscapes of \textit{C. elegans}, we curated CeNeuMorph ($n=27,117$), a multi-grained confocal dataset structured for phenotypic analysis at three levels of granularity (Fig.~\ref{fig1}b): image-level classification for global anomalies (break, arborization, bend); semantic segmentation for localized degenerative changes (bead); object detection for dendrite structures; and instance segmentation for specific neuronal bodies (CEP and ADE). 
A Basal Slowing Response (BSR) subset links these structural readouts to behavior.
Because neuronal pathology spans spatial scales, from few-pixel dendritic beading to whole-animal deformation, fixed-resolution patch embeddings impose a scale bottleneck that existing Masked Image Modeling (MIM) pre-training methods with fixed mask sizes struggle to resolve~\cite{beyer2023flexivit, he2022masked}. We thus developed a scale-adaptive masked-autoencoder framework that, during pre-training, jointly samples input resolutions and patch sizes under a fixed token budget, and at fine-tuning fuses two patch scales through a shared-weight encoder (Fig.~\ref{fig1}c,d; architecture detailed in Methods).

Evaluated on the CeNeuMorph benchmark, our scale-adaptive framework demonstrates improvements across multi-grained phenotypic tasks (Table~\ref{tab:model_comparison}). Under identical training data constraints and at a uniform input resolution of 224, our approach outperforms alternative pre-training methods-reaching 87.96\% accuracy in neuronal breakage detection compared to 86.40\% for the MAE baseline~\cite{he2022masked} and 78.27\% for SimMIM~\cite{xie2021simmim}.
Raising the pre-training input resolution to 480 lifted the breakage detection accuracy to 88.44\% ($+$1.19 percentage points over the strongest baseline, DINOv2 at 87.25\%). This indicates that our model exploits finer spatial details to extract structural semantics. Furthermore, our model surpasses generalist and biomedical foundation models (e.g., DINOv2~\cite{oquab2023dinov2}, MedSAM~\cite{ma2024segment}) that rely on large-scale pre-training corpora. Specifically, this advantage is evident in fine-grained tasks requiring precise boundary delineation: 78.26\% on dendritic beading semantic segmentation and 63.14\% and 70.10\% on ADE and CEP instance segmentation.

\begin{figure}[!htbp]
\centering
\includegraphics[width=1.0\textwidth, trim=0cm 14.5cm 0cm 0cm, clip]{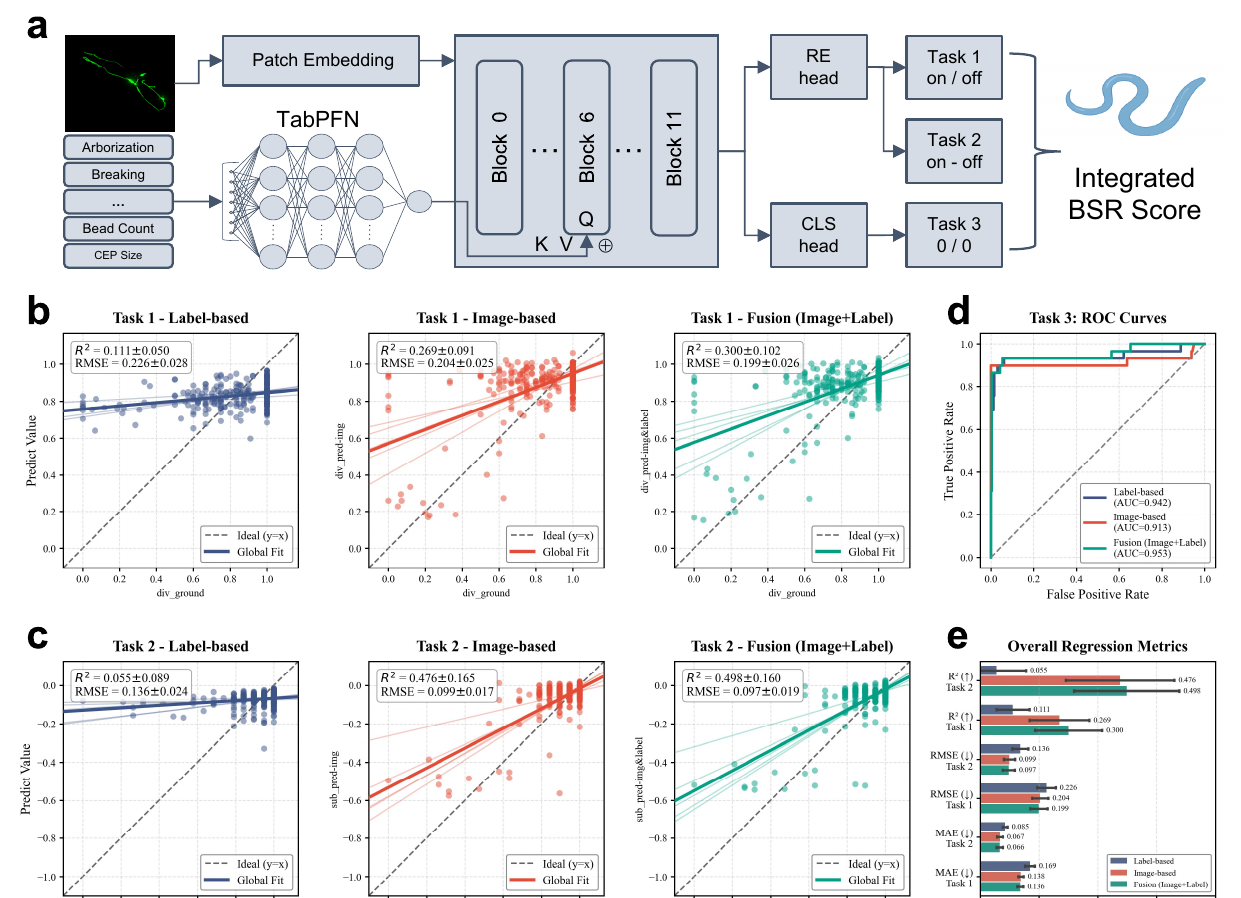}
\caption{\textbf{Quantitative assessment of behavioral trait prediction via morphological feature fusion.} 
(a) Schematic of the fusion model. TabPFN-embedded morphological labels integrate with visual patch embeddings via cross-attention at transformer Block 6 for downstream Regression (RE) and Classification (CLS). 
(b, c) Regression scatter plots for Task 1 ($f_{\mathrm{on}}/f_{\mathrm{off}}$) and Task 2 ($f_{\mathrm{on}}-f_{\mathrm{off}}$). Solid and faded lines denote global and individual 5-fold cross-validation fits, respectively. 
(d) ROC curves for Task 3 (0/0 status) classification, computed from pooled out-of-fold predictions across all 5 folds. 
(e) Summary of regression metrics ($R^2$, RMSE, MAE) for Tasks 1 and 2. Error bars represent the 5-fold cross-validation standard deviation, highlighting the Fusion approach's superior predictive accuracy.
}\label{fig2}
\end{figure}

\subsection{Fusing morphology with imaging predicts dopamine-dependent behavioral deficits}\label{subsec2}

To test whether structural readouts can forecast functional decline, we focused on the Basal Slowing Response (BSR)---a behavioral hallmark of dopaminergic health in~\textit{C. elegans} defined by the reduction in locomotive frequency upon encountering food ($f_{\mathrm{on}}$) compared to the off-food state ($f_{\mathrm{off}}$). 
Because absolute locomotion frequencies cannot be recovered from a static morphological snapshot, we framed prediction in terms of relative indices of dopaminergic modulation through three complementary tasks (defined in Methods): predicting the $f_{\mathrm{on}}/f_{\mathrm{off}}$ ratio (Task 1) and the $f_{\mathrm{on}}-f_{\mathrm{off}}$ difference (Task 2) to quantify the attenuation of the BSR, and identifying absolute immobility (Task 3: 0/0 status) as an indicator of severe functional failure. A dual-stream model fuses visual features from raw confocal images with TabPFN-encoded~\cite{hollmann2023tabpfn} morphological descriptors before a shared transformer backbone (Fig.~\ref{fig2}a; Methods). 

Comparative analysis (Fig.~\ref{fig2}b--e) demonstrates that integrating visual features with semantic morphological labels consistently optimizes the prediction of continuous behavioral traits. In Task 1 ($f_{\mathrm{on}}/f_{\mathrm{off}}$), the fusion model achieved an $R^2$ of 0.300, outperforming both image-only (0.269) and label-only (0.111) baselines. This superiority consistently extended to Task 2 ($f_{\mathrm{on}}-f_{\mathrm{off}}$), where the fusion approach yielded an $R^2$ of 0.498 and a minimum RMSE of 0.097, significantly surpassing single-stream configurations. For the binary classification of response status (Task 3), all models exhibited high discriminative power ($\mathrm{AUC} > 0.91$), yielding comparable performance across configurations (Fig.~\ref{fig2}d). This ceiling performance is biologically anticipated, as absolute immobility typically correlates with catastrophic and unambiguous morphological degradation. Overall, these results indicate that the fusion of structural semantic labels and raw visual data provides a robust framework for quantifying neurotoxic behavioral deficits.

\begin{figure}[htbp]
    \centering
    \includegraphics[width=1.0\textwidth, trim=0cm 2cm 0cm 0.5cm, clip]{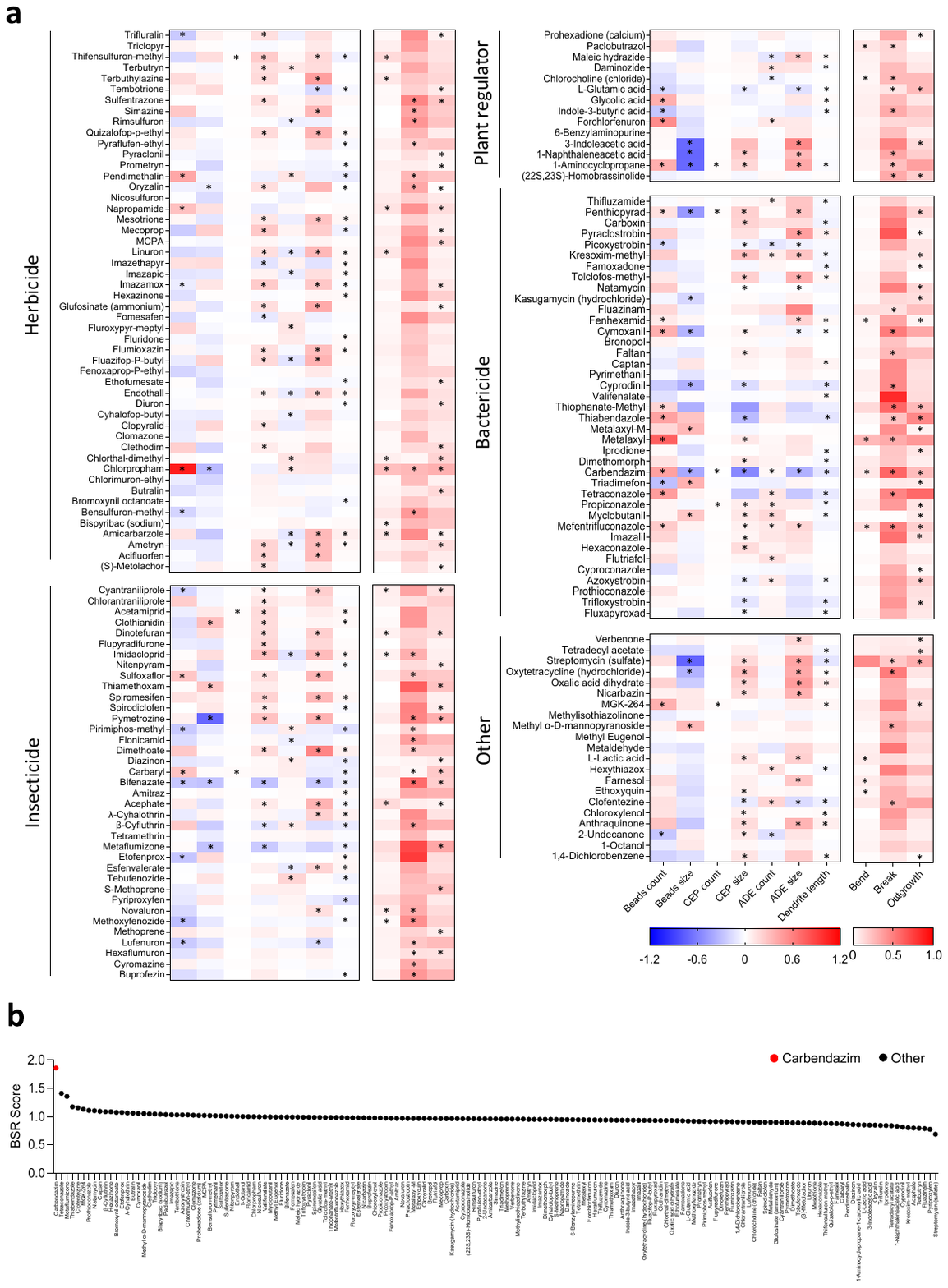}
    \caption{The damage identification model reveals that pesticides can cause damage to dopaminergic neurons in the head of \textit{C. elegans}. (a) Heatmap illustrating the effects of different pesticides (categorized as herbicides, insecticides, bactericides, plant regulators, and other compounds) on morphological indicators of dopaminergic neurons in Caenorhabditis elegans. Color scale represents change (blue: down-regulated; red: up-regulated; gray: no significant change). (b) BSR score ranking of the screened pesticides in descending order.}
    \label{fig3}
\end{figure}

\subsection{High-throughput screening links pesticides to dopaminergic injury}\label{subsec3}


We applied the injury-identification model to screen a panel of commonly used agrochemicals, comprising compounds epidemiologically linked to Parkinson's disease (PD) risk alongside many with no prior neurotoxicity reports~\cite{bib4,bib5}. 
Across the panel, pesticides induced distinct patterns of dopaminergic abnormalities in \textit{C. elegans}, that broadly tracked their functional class  (Fig.~\ref{fig3}a). Of the 180 compounds tested, 151 significantly altered at least one morphological indicator, with the affected indicators and the direction of change differing markedly across both pesticide classes and individual metrics (Fig.~\ref{fig3}a). Several hits were consistent with known mechanisms. Fluazinam and pyraclostrobin fell into the category of anticipated neurotoxicants, whose dopaminergic neuronal injury has previously been attributed to reactive oxygen species overproduction~\cite{bib7,bib8}.  Meanwhile, compounds such as chlorantraniliprole, previously characterized solely by acute toxic effects, also elicited signatures consistent with chronic dopaminergic neurotoxicity in our screening~\cite{bib11}. 

\begin{figure}[htbp]
    \centering
    \includegraphics[width=1.0\textwidth, trim=0cm 9cm 1cm 3cm, clip]{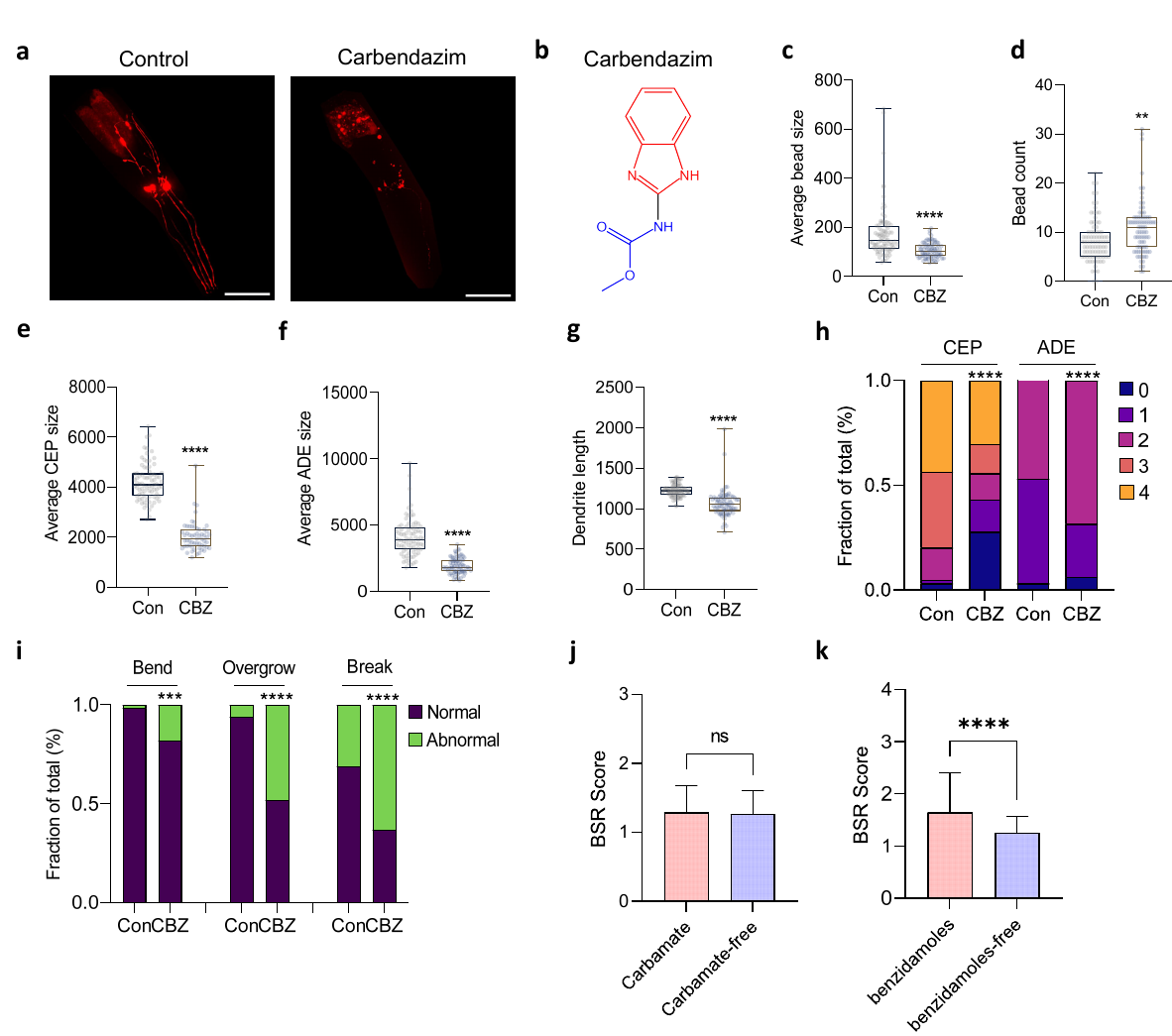}
    \caption{Benzimidazole ring is a structural determinant of dopaminergic neurotoxicity. (a) Representative micrograph of typical dopaminergic neuron morphology in carbendazim-exposed \textit{C. elegans}; scale bar = 50 $\mu$m. (b) Chemical structure of carbendazim, with the benzimidazole ring highlighted in red and the carbamate group in blue. (c–g) Statistics of average bead volume, bead number, soma volumes of CEP and ADE neurons, and dendritic length in carbendazim-exposed worms. (h) Soma counts of CEP and ADE neurons in carbendazim-exposed worms. (i) Incidence rates of abnormal bending, overgrowth, and breakage of CEP dendrites in carbendazim-exposed worms. (j) Statistical analysis of neurotoxicity associated with the carbamate moiety. (k) Statistical analysis of neurotoxicity associated with the benzimidazole ring.$^{\ast} p < 0.05$; $^{\ast\ast} p < 0.01$; $^{\ast\ast\ast} p < 0.001$; $^{\ast\ast\ast\ast} p < 0.0001$.}
    \label{fig4}
\end{figure}
As no single morphological readout captured the divergent injury phenotypes, we used the integrated BSR score as a unified quantitative measure of overall neurotoxic burden (Fig.~\ref{fig3}b). Strikingly, fungicide carbendazim exhibited the strongest neurotoxicity among all tested agrochemicals, with a BSR score of 1.85---85\% higher than vehicle control levels and 32\% above the second-ranked tetraconazole (BSR = 1.40). Prior functional studies have linked carbendazim to both neuronal apoptosis and the disruption of protein degradation pathways in dopaminergic neurons~\cite{bib12,bib13,bib14}. We therefore performed detailed morphological profiling of carbendazim in the subsequent section.

\begin{figure}[htbp]
    \centering
    \includegraphics[width=1.0\textwidth, trim=0cm 8.5cm 0cm 4cm, clip]{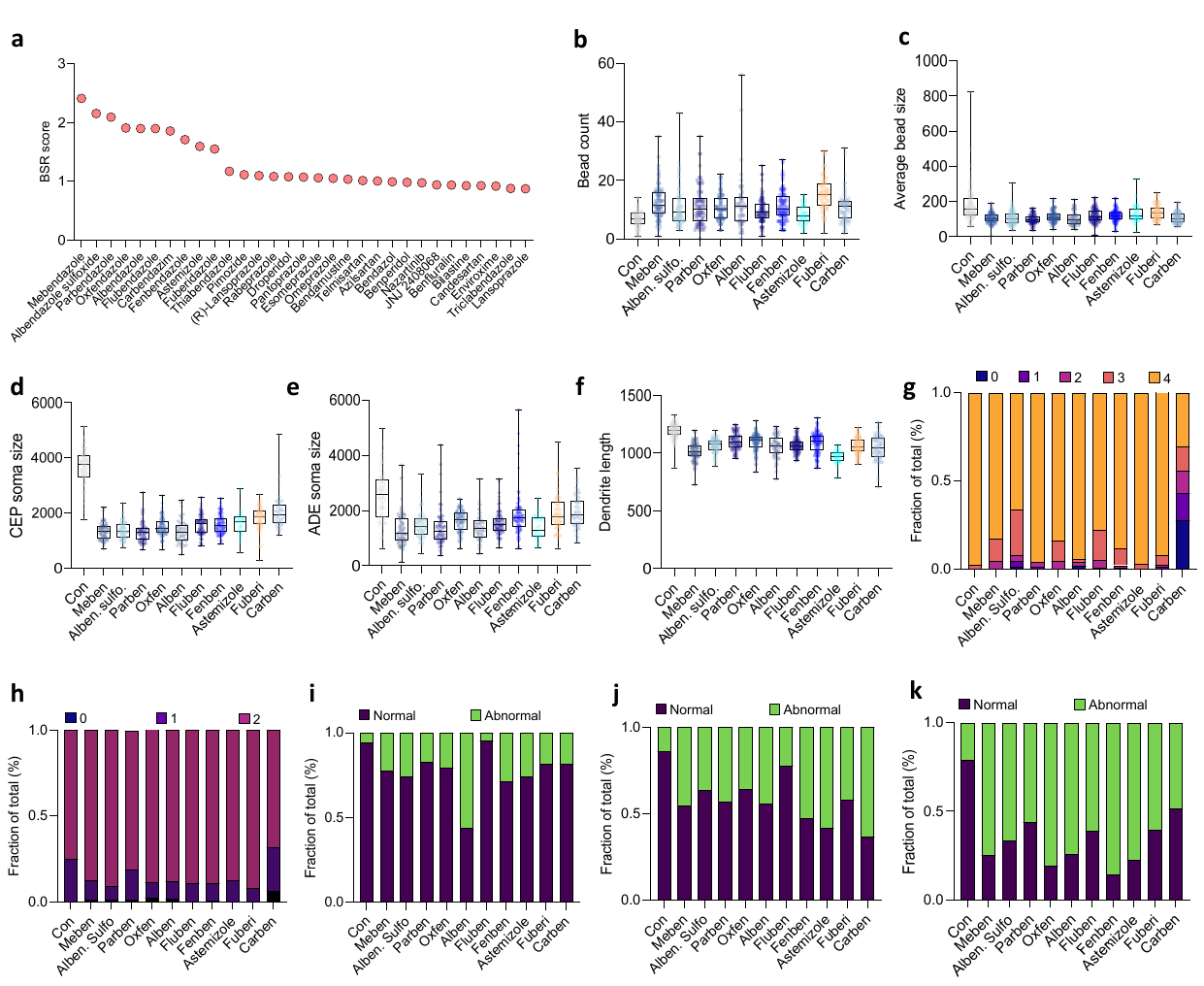}
    \caption{Benzimidazole pesticides and drugs impair dopaminergic neuron morphology in \textit{C. elegans} and show high similarity in neuronal morphology. (a) BSR score ranking of benzimidazole-containing compounds in descending order. Panels (b–k) only include benzimidazole compounds with a BSR score $> 1.5$. (b–j) Box plots quantifying dopaminergic neuron morphological parameters: (b) bead count, (c) average bead size, (d) CEP soma size, (e) ADE soma size, and (f) dendrite length, in benzidamole drugs. (f–j) Stacked bar charts showing the fraction of worms with normal/abnormal phenotypes: (g) CEP neuron count, (h) ADE neuron count, (i) abnormal bend, (j) break, and (k) outgrowth, across control and benzimidazole-treated groups.
}
    \label{fig5}
\end{figure}

\subsection{The benzimidazole ring is a structural determinant of dopaminergic neuroroxicity}\label{subsec4}

We next examined carbendazim-triggered morphological defects to resolve its distinct effects on dopaminergic neuronal morphology. Further microscopic examination of dopaminergic neurons following carbendazim exposure revealed their near-complete ablation (Fig.~\ref{fig4}a). The morphological pathology of dopaminergic neurons caused by this fungicide was characterized by increased bead counts, reduced numbers and shrinkage of neuronal soma, shortened dendritic lengths, and elevated incidences of abnormal curvature, hyperplasia, and breakage in CEP dendrites (Fig.~\ref{fig4}c-i), effects that significantly distinguished it from other severely damaging pesticides (Fig.~\ref{fig_A3}). Carbendazim is chemically simple, consisting of a benzimidazole ring and a carbamate side chain (Fig.~\ref{fig4}b). This prompted us to ask which structural moiety underlies such potent neurotoxicity. Given the limited number of benzimidazole compounds (only two compounds) in our initial library, we first analyzed the contribution of the carbamate group, with eight carbamate-containing agrochemicals present within our compound panel. Our results showed that the presence or absence of a carbamate moiety did not affect the dopaminergic neuronal injury (Fig.~\ref{fig4}j). To verify the role of the benzimidazole ring, we subsequently examined 29 commercially available benzimidazole-containing pharmaceutical chemicals. These experiments confirmed that these compounds harboring the benzimidazole ring exhibited a pronounced preference for dopaminergic neurotoxicity (Fig.~\ref{fig4}k), among these compounds, ten drugs yielded BSR scores exceeding the vehicle control by over 50\% (BSR $> 1.5$). This subset was dominated by benzimidazole insecticides and fungicides, indicating that these agrochemicals, rather than clinical benzimidazole therapeutics, serve as the primary triggers of dopaminergic neuronal damage (Fig.~\ref{fig5}a). Intriguingly, these severely neurotoxic compounds cause prominent morphological damage marked by near-complete loss of dopaminergic neurons relative to vehicle controls (Fig.~\ref{fig5}b-k). By contrast, clinical benzimidazole compounds did not show prominent morphological alterations (Fig.~\ref{fig_A4}). Taken together, these findings establish the benzimidazole scaffold as a previously overlooked structural determinant of dopaminergic vulnerability. Furthermore, these phenotypic observations underscore the ability of our screening approach to resolve structure-dependent neurotoxic liabilities within a given chemical class. 

\section{Discussion}\label{sec3}

Quantifying neurotoxicity has long been caught between two disconnected readouts: morphological imaging, which reports structural change, and behavioral assays, which report functional consequence. Bridging them is simultaneously a computational problem---resolving lesions that span spatial scales within a tractable model---and a biological one---establishing that structure predicts function. Here we addressed both. We developed a scale-adaptive vision model that detects subtle morphological alterations in the cephalic dopaminergic neurons of \textit{C. elegans} and links them to behavioral responses, and we subsequently applied it to screen pesticides, some of which have been associated with Parkinson's disease risk. 


Resolving hierarchical neurodegenerative pathology in sparse microscopy is computationally limited by standard Masked Autoencoders (MAE), which tend to reconstruct dark backgrounds over minute biological signals and impose a severe spatial trade-off where small patches incur prohibitive fine-tuning costs~\cite{he2022masked}.
We address these limitations by developing an improved scale-adaptive Masked Image Modeling (MIM) strategy combined with FlexiViT~\cite{he2022masked, beyer2023flexivit}. Randomizing resolutions and patch sizes under a constant token budget decouples semantic learning from rigid spatial grids. This mechanism serves as a biological inductive bias, forcing the encoder to simultaneously ``zoom in'' on localized lesions and ``zoom out'' for global anatomical context. By explicitly targeting confocal imaging properties, this scale-aware architecture efficiently outperforms the brute-force data scaling of generalist foundation models in specialized biomedical domains. 
An ablation study (Table~\ref{tab:table_s1}) confirms that dynamic scaling and length-constrained masking establish a scale-invariant foundation, yielding systematic improvements across all multi-grained morphological tasks. Building upon this, multi-scale fusion captures the high-resolution spatial details required for dense segmentation, seamlessly integrating localized micro-lesions with macroscopic deformations without additional parameter overhead.
Alongside the open CeNeuMorph benchmark, this scale-adaptive strategy provides a reusable, readily transferable foundation for other biomedical imaging domains.

Our framework operationalizes a morphology-to-behavior mapping that conventional \textit{C. elegans} neurotoxicity assessments have left implicit, treating structural degeneration and functional decline as disconnected readouts: manual morphological scoring fails to capture holistic neuronal health~\cite{bib3,bib15}, while behavioral output lacks mechanistic specificity and throughput~\cite{bib1,bib16,bib17}. Rather than the standard late-fusion ensembles that dilute localized feature variances, the architecture adopts an intermediate-fusion topology (Figure~\ref{fig_A2}). By embedding TabPFN-encoded structural priors into the Vision Transformer's mid-level semantic manifold~\cite{hollmann2023tabpfn}, the model seamlessly fuses visual and morphological features to drive biologically meaningful multi-task predictions (Tasks 1--3). By accurately tracking continuous functional decline (Task 1: $f_{on}/f_{off}$; Task 2: $f_{on}-f_{off}$), the model demonstrates that explicit pathological priors (semantic stream) and subtle structural degradations (visual stream) are highly complementary for predicting progressive dopaminergic attenuation. Furthermore, its precise discrimination of absolute immobility (Task 3: $0/0$ status) strictly mirrors biological reality: catastrophic functional collapse inherently stems from gross morphological breakdown. Ultimately, establishing this structural-functional link successfully facilitated the high-throughput neurotoxicity screening of pesticide compounds in \textit{C. elegans}.

Dopaminergic neurons in~\textit{C. elegans}, with their conserved morphology and stable physiology, provide a well-established model for neurodegeneration and Parkinson's disease~\cite{bib18,bib62}, a natural substrate for prioritizing environmental chemicals by neurotoxic risk. 
Applied across a panel of commonly used agrochemicals, we revealed that pesticides produce distinct patterns of dopaminergic abnormality that broadly track functional class: of 180 compounds screened, 151 perturbed at least one morphological indicator, with the affected features and the direction of change differing markedly across both class and metric. 
Our findings highlight that neuronal degeneration cannot be fully characterized by cell loss alone~\cite{bib18,bib46}. Fine-grained morphological features, including dendritic beading, arborization abnormalities, and soma remodeling, provide sensitive indicators of neuronal injury that may precede overt neurodegeneration~\cite{bib63,bib64}. By combining multi-grained morphological phenotyping with behavioral readouts, the platform offers enhanced sensitivity for early neurotoxicity detection and enables systematic assessment of structure–function relationships. This application successfully identified both previously recognized pesticides associated with Parkinson's disease~\cite{bib47,bib48,bib7, bib21}, confirming the sensitivity and discovery potential of our screening platform. Moreover, unlike target-based screens that can miss off-target signals, morphological assessment captures unbiased phenotypic alterations~\cite{bib6,bib67}. Our AI-driven phenotypic screening thus offers a novel validation strategy to uncover unanticipated neurotoxicity.

A particularly notable finding was the recurrent enrichment of benzimidazole-containing compounds among the strongest predictors of behavioral impairment and dopaminergic neuronal damage. Although benzimidazole derivatives are widely used in agriculture, veterinary medicine, and clinical therapeutics, their potential effects on dopaminergic neurons have received limited attention~\cite{bib22,bib49,bib50}. Our results demonstrate that multiple benzimidazole-based pesticides and veterinary anthelmintics induce highly similar patterns of neuronal injury, identifying the benzimidazole moiety as a previously underrecognized structural feature associated with dopaminergic neurotoxicity. 
Notably, carbamate pesticides elicited only minimal dopaminergic neurotoxic signals in our screen. This is mechanistically coherent with their established toxicological profile: carbamates reversibly inhibit acetylcholinesterase (AChE) with spontaneous decarbamoylation occurring within approximately 30 minutes, and are rapidly metabolized and excreted within 24 hours without tissue accumulation. Even if cholinergic dysregulation could indirectly influence dopaminergic circuits, the transient nature of AChE inhibition and swift clearance preclude the sustained subcellular damage required for progressive neurodegeneration, accounting for the weak phenotypes observed in our assays~\cite{bib65,bib66}.
At the same time, the substantially lower neurotoxicity observed for clinically used benzimidazole drugs suggests that, although the benzimidazole ring may represent a necessary structural determinant of toxicity, its neurotoxic potential is profoundly influenced by substituent chemistry and other molecular characteristics. These findings indicate that appropriate structural optimization can effectively mitigate neurotoxicity, highlighting a critical safe-by-design strategy for the development and clinical application of benzimidazole-based compounds. These observations underscore the importance of evaluating neurotoxicity at the level of chemical classes rather than individual compounds and highlight the potential value of structure-guided toxicological assessment. More broadly, our study establishes an integrated framework for large-scale neurotoxicant discovery and provides a foundation for future investigations into the molecular mechanisms linking environmental chemical exposure to dopaminergic neurodegeneration.

While our framework establishes a robust paradigm for automated neurotoxicity assessment, it also illuminates several distinct avenues for future advancement. First, the current reliance on 2D maximum intensity projections inherently compresses intricate spatial topography. Transitioning to native 3D volumetric pre-training architectures, coupled with scaling the training corpora to encompass all 302 neurons of \textit{C. elegans}, will exploit $z$-axis depth context and pave the way for a whole-nervous-system foundation model. Second, to refine the morphology-behavior mapping, future models should move beyond static structural correlations and discrete functional readouts. The current BSR assessment relies on only two locomotive values. Directly linking neuronal morphology with continuous behavioral videos, alongside broader contexts such as age, genetic strains, and environmental factors, will yield a more holistic model of neurodegeneration. Ultimately, the broader implications of this work extend in two directions: methodologically, the scale-adaptive pre-training strategy provides a transferable blueprint for other sparse biomedical imaging domains; biologically, the high-throughput screening paradigm can be readily expanded to broader chemical libraries and adapted to other genetically tractable model organisms (e.g., zebrafish or \textit{Drosophila}), significantly accelerating environmental risk evaluation and neuroprotective drug discovery.


\section{Methods}\label{sec4}

\subsection{Experimental workflow and data acquisition}\label{subsec1}

To construct a robust and highly standardized dataset for neurotoxicity screening, the \textit{in vivo} data acquisition pipeline was structured into four sequential phases: compound library curation, controlled drug exposure, high-resolution phenotypic imaging, and behavioral metric quantification. All cultures and experimental environments were rigorously maintained at a constant 20 $^\circ$C, utilizing the N2 Bristol strain to establish wild-type baseline parameters.

\textbf{Compound Library Filtering.} We started with a commercial compound library (MCE) consisting of 309 EPA-approved pesticides and agriculturally relevant molecules. A stringent filtering strategy was then implemented: compounds with known associations with Parkinson's disease or confirmed nematicidal activities were excluded, aiming to identify novel agents with potential dopaminergic neurotoxicity. After filtration, a refined set of 180 compounds covering multiple functional categories (herbicides, bactericides, fungicides, plant growth regulators and acaricides) was obtained for subsequent screening. 

\textbf{Drug Exposure.} Stock solutions of all compounds were prepared in DMSO at 10 mM. Before experiments, each compound was diluted in ddH$_2$O supplemented with 5-fluoro-2$^{\prime}$-deoxyuridine (Aladdin). The mixture was evenly spread onto Nematode Growth Medium (NGM) plates to reach final concentrations of 100 $\mu$M (compound) and 50 $\mu$M (5-fluoro-2$^{\prime}$-deoxyuridine). Plates were dried overnight, then L4 stage hermaphroditic \textit{C. elegans} (N2 wild-type and \textit{YmIs20} [P\textit{dat-1::SL2::dsRed} $+$ P\textit{lin-44::mCherry}]) were transferred onto the plates. Worm survival was counted daily.

\textbf{Phenotype Acquisition.} On day seven of exposure, worms were anesthetized with 10 mM NaN$_3$ on a glass slide coated with a thin layer of 3\% agarose (Yeasen). Confocal images of head dopaminergic neurons were acquired using a $60\times$ oil objective on a live-cell spinning disk confocal microscope (OLYMPUS Spin10 (CSU-W1)) with the following parameters: exposure time 100 ms, Z-stack step size 1 $\mu$m, laser intensity 50\%, and resolution $2304 \times 2304$ pixels.

\textbf{Behavioral Assays.} For behavioral assays, seven-day-old worms were used. The Basal Slowing Response (BSR) was tested by washing worms twice with M9 buffer, transferring them to unseeded NGM plates, and counting body bends for 20 s (on-food value); after 5 min, body bends were counted again (off-food value). All cultures and experiments were maintained at 20 $^\circ$C, with N2 Bristol as the wild-type control. Locomotion modulation in response to food stimuli was assessed based on previous studies~\cite{bib24,bib25}.




\subsection{Datasets and data processing}\label{sebsec5}
Raw volumetric data of \textit{C. elegans} head dopaminergic neurons were acquired via live-cell spinning disk confocal microscopy at a spatial resolution of $2304 \times 2304$ pixels and converted into 2D representations using Z-axis maximum intensity projection. Ground-truth annotations were performed blindly through random image renaming, followed by independent cross-validation by two neurotoxicology domain experts. For global anomalies, dendrite break, outgrowth, and abnormal bend were annotated as binary variables (1/0 indicating presence or absence), and dendrite length was measured via the minimal bounding box after rotating neurons to a horizontal/vertical orientation. Pixel-level annotations for dendritic beads and soma structures were executed using LabelMe (v5.5.0) to generate training JSON files~\cite{bib23}. Punctate dendritic beading was strictly defined as 0.5--4 $\mu$m round structures, explicitly excluding exophers and non-specific aggregates.

For data preprocessing and augmentation, images within the classification subsets underwent histogram equalization to standardize contrast prior to modeling. To optimize the instance segmentation of neuronal somata, an initial set of 250 high-resolution images was processed using Slicing Aided Hyper Inference (SAHI)~\cite{akyon2022sahi} for tiled cropping, yielding a refined subset of 1,938 localized spatial samples. The resulting stratified subsets comprised dendritic arborization ($n=2,807$), neuronal breakage ($n=2,807$), axonal bending ($n=498$), dendritic beading ($n=3,906$), dendrite object detection ($n=250$), CEP/ADE soma instance segmentation ($n=1,938$) and a Basal Slowing Response (BSR) subset ($n=288$). 

Across all enumerated subsets, we first reserved a fixed 15\% of the data as an independent test set. The remaining 85\% was used for 5-fold cross-validation to allocate training and validation sets. The final metrics in Table~\ref{tab:model_comparison} represent the mean and standard deviation of these five models evaluated on the fixed test set. As a single exception, the smaller BSR cohort was evaluated via direct 5-fold cross-validation without a reserved test set, thereby maximizing statistical power and avoiding the high variance inherent to small sample sizes.

\subsection{Scale-adaptive MIM pre-training}\label{sebsec6}

Biological structures exhibit intrinsic multi-scale characteristics, ranging from localized neuronal defects to global geometric deformations. Standard Vision Transformers (ViTs) utilize fixed-resolution patch embeddings (typically $16 \times 16$)~\cite{dosovitskiy2020image}, imposing a scale bottleneck that limits the simultaneous resolution of fine and coarse features. To address this, we developed a scale-adaptive Masked Image Modeling (MIM) framework that integrates a flexible patch embedding module~\cite{beyer2023flexivit}, effectively decoupling patch size from rigid grid constraints.

During pre-training, input images undergo stochastic scaling to a randomly sampled target resolution (e.g., $[240, 480]$ for the $480 \times 480$ pipeline, or $[112, 224]$ for the $224 \times 224$ pipeline), alongside a randomly sampled patch size ($p \in [4, 32]$ pixels). Images are bicubically interpolated, and patch embedding weights are dynamically resized to match the specified dimensions. This projection yields a variable-length 1D token sequence ranging from 225 to 900 tokens. To maintain a unified computational footprint, we implement a sequence-length-constrained masking strategy rather than a standard fixed masking ratio~\cite{he2022masked}. The embedded sequence is shuffled and strictly truncated to exactly 180 visible tokens ($L_{keep} = 180$). This constraint forces the encoder to reconstruct global morphological semantics from a constant informational budget, irrespective of the initial input scale.

The architecture comprises a 12-block ViT encoder (embedding dimension 768, 12 attention heads) and an 8-block decoder (dimension 512, 16 attention heads). The unmasked tokens are processed by the encoder, concatenated with learnable mask tokens, and routed through the decoder to reconstruct dense pixel values at the original macroscopic resolution(e.g., $480 \times 480$ or $224 \times 224$). The model is optimized using a per-patch normalized mean squared error (MSE) loss. 
Models were initialized with standard MAE weights~\cite{he2022masked} and pre-trained for 800 epochs using the AdamW optimizer ($\beta_1=0.9, \beta_2=0.95$, weight decay 0.05). Training utilized a base learning rate of $1 \times 10^{-3}$, a 40-epoch linear warmup, and an effective batch size of 64, distributed across four NVIDIA A100 GPUs.

\subsection{Adaptation to downstream tasks}\label{sebsec7}

For downstream fine-tuning, the pre-training decoder is discarded. To address the multi-grained nature of the morphological tasks, the Vision Transformer backbone is adapted using a parameter-sharing dual-scale processing strategy. Parallel streams process inputs at fine ($p=16$) and coarse ($p=32$) resolutions using identical model weights. This approach maximizes phenotypic precision through increased computation without introducing additional parameter overhead. Cross-scale information complementation is facilitated by iteratively exchanging \texttt{[CLS]} tokens between the two branches after each transformer block.

For image-level classification of global anomalies (neuronal breakage, dendritic arborization, and axonal bending) at $480 \times 480$ resolution, standard \texttt{[CLS]} classification is bypassed. Instead, spatial tokens from the coarse scale are modulated by a learnable spatial weight parameter and subsequently concatenated with the fine-scale token sequence. The fused multi-scale representation is then subjected to global average pooling (GAP). This aggregation captures both local textural anomalies (e.g., small breaks) and global shape deformations (e.g., deep bends), yielding robust translation-invariant representations for task-specific prediction.

For dense prediction tasks requiring precise spatial localization---semantic segmentation ($480 \times 480$), object detection ($640 \times 640$), and instance segmentation ($640 \times 640$)---the backbone extracts dual-resolution spatial feature maps at $1/16$ and $1/32$ scales. Following the ViTDet paradigm, independent hierarchical feature pyramids constructed for each branch are aligned via progressive upsampling and lateral convolutions, and subsequently fused~\cite{li2022exploring}. The resulting unified representations are directly routed to OpenMMLab pipelines: MMSegmentation~\cite{mmseg2020} for semantic segmentation of dendritic beading, and MMDetection~\cite{mmdetection} for dendrite object detection ($n=250$) alongside CEP/ADE soma instance segmentation ($n=1,938$).

\subsection{Modeling the morphology-behavior relationship}
To quantify the functional impact of structural degradation, we formulated a multi-task learning framework centered on the Basal Slowing Response (BSR), a behavioral hallmark of dopaminergic health in \textit{C. elegans} defined by the reduction in locomotive frequency upon encountering food ($f_{\mathrm{on}}$) relative to the off-food state ($f_{\mathrm{off}}$). Because static confocal snapshots inherently underdetermine absolute kinematic frequencies, the functional assessment is parameterized into three relative indices of dopaminergic modulation: predicting the $f_{\mathrm{on}}/f_{\mathrm{off}}$ ratio (Task 1) and the $f_{\mathrm{on}}-f_{\mathrm{off}}$ difference (Task 2) to quantify BSR attenuation, and classifying absolute immobility ($0/0$ status, Task 3) as a binary indicator of severe functional collapse.

We design a dual-stream for multimodal feature integration. The visual stream extracts high-dimensional spatial patch embeddings from raw Z-projected micrographs. Concurrently, the semantic stream explicitly encodes a comprehensive suite of phenotypic descriptors using a TabPFN encoder~\cite{hollmann2023tabpfn}. These descriptors encompass the binary downstream classification outcomes derived from the fine-tuned models---specifically neuronal breaking, arborization, and abnormal bending---alongside quantified morphological metrics, including bead count, average bead size, dendrite length, and the respective soma counts and average sizes of CEP and ADE neurons.

To integrate these representations, we employ an intermediate cross-attention fusion module at the sixth block of the 12-block transformer backbone. Specifically, the mid-level spatial patch embeddings act as queries to selectively attend to the TabPFN-encoded semantic vectors (serving as keys and values). A zero-initialized gating mechanism ensures stable residual integration, smoothly conditioning the visual representations on explicit morphological priors without disrupting the pre-trained weights. Terminal features are subsequently routed to a Regression (RE) head for continuous BSR indicators and a Classification (CLS) head for binary response status. The independently normalized predictions are ultimately aggregated into a composite BSR injury score via a weighted summation (1:1:2 for Tasks 1, 2, and 3, respectively), assigning double weight to Task 3 to explicitly account for the catastrophic neurodegenerative damage associated with absolute functional failure.

\subsection{Baseline and foundation-model implementation}\label{sebsec9}

We benchmarked our framework against two categories of state-of-the-art architectures: self-supervised pre-training baselines and large-scale vision foundation models.
The pre-training baselines encompassed representative contrastive and masked image modeling (MIM) paradigms: MoCo v3~\cite{chen2021mocov3}, MAE~\cite{he2022masked}, MSN~\cite{assran2022msn}, SimMIM~\cite{xie2021simmim}, MixMAE~\cite{liu2023mixmae}, and HiViT~\cite{zhang2023hivit}. To ensure standardized evaluation, these models were initialized with ImageNet-1K weights and pre-trained on our 23,150 \textit{C. elegans} confocal images using their original architectural and hyperparameter configurations.
To assess domain transferability, we additionally evaluated generalist (EVA-02~\cite{fang2024eva}, DINOv2~\cite{oquab2024dinov2}, SigLIP2~\cite{tschannen2025siglip}) and biomedical (MedSAM~\cite{ma2024medsam}, BiomedCLIP~\cite{zhang2024biomedclip}, and MedImageInsight~\cite{codella2024medimageinsight}) foundation models. 
Following the respective pre-training or representation extraction phases, all models underwent an identical downstream fine-tuning protocol. We transferred the encoded representations and subjected them to the same downstream pipelines for image-level classification, dense semantic and instance segmentation, and object detection tasks, allowing for a standardized assessment of feature robustness and phenotypic resolution.


\section*{Data availability}
The multi-grained \textit{C. elegans} confocal microscopy dataset (CeNeuMorph) supporting the findings of this study has been deposited in a public repository \url{https://doi.org/10.5281/zenodo.21534168} and is horizontally indexed for academic use. Fully annotated image subsets for downstream benchmarking---including morphological classification, semantic segmentation of dendritic beading, and instance-level neuronal structure analysis---are available within the repository. All other intermediate data or raw microscopy frames that support the plots and statistical evaluations within this paper are available upon reasonable request.

\section*{Code availability}
The complete pipeline for our scale-adaptive self-supervised learning framework, including the scale-adaptive MIM pre-training code and downstream task adaptation scripts (integrated with \textit{MMDetection} and \textit{MMSegmentation}), has been made publicly available on GitHub at \url{https://github.com/ZJUMAI/CelegansNeuro-Vision}. The core architecture is implemented in PyTorch, and a comprehensive instruction manual, along with pre-trained ViT-B backbone weights, is provided to ensure full reproducibility of the experimental benchmarks.

\backmatter



\bigskip

\bibliography{sn-bibliography}

@article{bib1,
  author		= "Morton, K.S. and George, A.J. and Meyer, J.N.",
  title			= "Complex I superoxide anion production is necessary and sufficient for complex I inhibitor-induced dopaminergic neurodegeneration in Caenorhabditis elegans.",
  journal		= "Redox Biol",
  volume		= "81",
  pages			= "103538",
  year			= "2025"
}

@article{bib3,
  author		= "Sawin, E.R. and Ranganathan, R. and Horvitz, H.R.",
  title			= "C. elegans locomotory rate is modulated by the environment through a dopaminergic pathway and by experience through a serotonergic pathway.",
  journal		= "Neuron",
  volume		= "26",
    number  ="3",
  pages			= "619-31",
  year			= "2000"
}

@article{bib4,
  author		= "Polymeropoulos, M.H. and Lavedan, C. and Leroy, E.",
  title			= "Mutation in the alpha-synuclein gene identified in families with Parkinson's disease.",
  journal		= "Science",
  volume		= "276",
    number  ="5321",
  pages			= "2045-71",
  year			= "1997"
}

@article{bib5,
  author		= "Dorsey, E.R. and DeMiranda, B.R. and Hussain, S.",
  title			= "Environmental toxicants and Parkinson's disease: recent evidence, risks, and prevention opportunities.",
  journal		= "Lancet Neurol",
  volume		= "24",
    number  ="11",
  pages			= "976-986",
  year			= "2025"
}

@article{bib6,
  author		= "Paul, K.C. and Krolewski, R.C. and Lucumi, M. E.",
  title			= "A pesticide and iPSC dopaminergic neuron screen identifies and classifies Parkinson-relevant pesticides.",
  journal		= "Nat Commun",
  volume		= "14",
    number  ="1",
  pages			= "2803",
  year			= "2023"
}

@article{bib7,
  author		= "Lee, J.E. and Kang， J.S. and Ki, Y.W.",
  title			= "Fluazinam targets mitochondrial complex I to induce reactive oxygen species-dependent cytotoxicity in SH-SY5Y cells",
  journal		= "Neurochem Int",
  volume		= "60",
    number  ="8",
  pages			= "773-81",
  year			= "2012"
}

@article{bib8,
  author		= "Huayta, J. and Meyer, J.N.",
  title			= "Inhibition of Mitochondrial Complex III Causes Dopaminergic Neurodegeneration by Redox Stress in \textit{Caenorhabditis elegans}",
  journal		= "bioRxiv",
  volume		= "23",
  pages			= "2025.10.21.683798",
  year			= "2025"
}

@article{bib11,
  author		= "Kimura, M. and Shoda, A. and Murata, M.",
  title			= "Neurotoxicity and behavioral disorders induced in mice by acute exposure to the diamide insecticide chlorantraniliprole",
  journal		= "J Vet Med Sci",
  volume		= "85",
number="4",
  pages			= "497-506",
  year			= "2023"
}

@article{bib12,
  author		= "Fitzmaurice, A.G. Rhodes, S.L. and Lulla, A.",
  title			= "Aldehyde dehydrogenase inhibition as a pathogenic mechanism in Parkinson disease.",
  journal		= "Proc Natl Acad Sci U S A",
  volume		= "110",
  number="2",
  pages			= "636-41",
  year			= "2013"
}

@article{bib13,
  author		= "Tufi, S. and Wassenaar, P.N. and Osorio, V.H.",
  title			= "Pesticide Mixture Toxicity in Surface Water Extracts in Snails (Lymnaea stagnalis) by an in Vitro Acetylcholinesterase Inhibition Assay and Metabolomics.",
  journal		= "Environ Sci Technol",
  volume		= "50",
number="7",
  pages			= "3937-44",
  year			= "2016"
}

@article{bib14,
  author		= "Ebedy, Y.A. and Hassanen, E.I. and Hussien, A.M.",
  title			= "Neurobehavioral Toxicity Induced by Carbendazim in Rats and the Role of iNOS, Cox-2, and NF-$\kappa$B Signalling Pathway.",
  journal		= "Neurochem Res",
  volume		= "47",
number="7",
  pages			= "1956-1971",
  year			= "2022"
}

@article{bib15,
  author		= "Tanimoto, Y. and Zhang, Y.G. and Fei, X.",
  title			= "In actio optophysiological analyses reveal functional diversification of dopaminergic neurons in the nematode C. elegans.",
  journal		= "Sci Rep",
  volume		= "6",
  pages			= "26297",
  year			= "2016"
}

@article{bib16,
  author		= "Martinez-Finley, E.J. and Chakraborty, S. and Caito, S.",
  title			= "\textit{C. elegans} and Neurodegeneration \textit{In Caenorhabditis Elegans: Anatomy, Life Cycles and Biological Functions}",
  journal		= "Adv Med Biol",
  volume		= "44",
  pages			= "1-46",
  year			= "2012"
}

@article{bib17,
  author		= "Clark, A.S. and Kalmanson, Z. and Morton, K.",
  title			= "An unbiased, automated platform for scoring dopaminergic neurodegeneration in C. elegans.",
  journal		= "PLoS One",
  volume		= "18",
number="7",
  pages			= "e0281797",
  year			= "2023",
}

@article{bib18,
  author		= "Chikka, M.R. and Anbalagan, C. and Dvorak, K.",
  title			= "The Mitochondria-Regulated Immune Pathway Activated in the C. elegans Intestine Is Neuroprotective.",
  journal		= "Cell Rep",
  volume		= "16",
number="9",
  pages			= "2399-414",
  year			= "2016",
}

@article{bib21,
  author		= "Sanchez, C.L. and Souders, C.L. and Pena-Delgado, C.J.",
  title			= "Neurotoxicity assessment of triazole fungicides on mitochondrial oxidative respiration and lipids in differentiated human SH-SY5Y neuroblastoma cells.",
  journal		= "Neurotoxicology",
  volume		= "80",
  pages			= "76-78",
  year			= "2020",
}

@article{bib22,
  author		= "Keri, R.S. and Hiremathad, A. and Budagumpi, S.",
  title			= "Comprehensive Review in Current Developments of Benzimidazole-Based Medicinal Chemistry.",
  journal		= "Chem Biol Drug Des",
  volume		= "86",
number="1",
  pages			= "19-65",
  year			= "2015",
}

@article{bib23,
  author		= "Melentijevic, I. and Toth, M.L. and Arnold, M.L.",
  title			= "C. elegans neurons jettison protein aggregates and mitochondria under neurotoxic stress.",
  journal		= "Nature",
  volume		= "542",
number="7641",
  pages			= "367-371",
  year			= "2017",
}

@article{bib24,
  author		= "Pau, Friedman, K. and Gagne, M. and Loo, L.H.",
  title			= "Utility of In Vitro Bioactivity as a Lower Bound Estimate of In Vivo Adverse Effect Levels and in Risk-Based Prioritization.",
  journal		= "Toxicol Sci",
  volume		= "173",
number="1",
  pages			= "202-225",
  year			= "2020",
}

@article{bib25,
  author		= "Wang, S. and Jiang, Y. and Yang, A.",
  title			= "The Expanding Burden of Neurodegenerative Diseases: An Unmet Medical and Social Need.",
  journal		= "Aging Dis",
  volume		= "16",
number="5",
  pages			= "2937-2952",
  year			= "2024",
}

@article{bib46,
      author		= "Nass, R. and Blakely, R.D.",
  title			= "The Caenorhabditis elegans dopaminergic system: opportunities for insights into dopamine transport and neurodegeneration",
  journal		= "Annu Rev Pharmacol Toxicol",
  volume		= "43",
  pages			= "521-44",
  year			= "2003",
}

@article{bib47,
      author		= "van der Mark M. and Brouwer, M. and Kromhout, H.",
  title			= "Is pesticide use related to Parkinson disease? Some clues to heterogeneity in study results",
  journal		= "Environ Health Perspect",
  volume		= "120",
  pages			= "340-7",
  year			= "2012",
}

@article{bib48,
      author		= "Gunnarsson, L.G. and Bodin, L.",
  title			= "Occupational Exposures and Neurodegenerative Diseases-A Systematic Literature Review and Meta-Analyses",
  journal		= "Int J Environ Res Public Health",
  volume		= "16",
  pages			= "337",
  year			= "2019",
}

@article{bib49,
      author		= "Bai, S. and Zhang, M. and Tang, S.",
  title			= "Research Progress on Benzimidazole Fungicides: A Review",
  journal		= "Molecules",
  volume		= "29",
  pages			= "1218",
  year			= "2024",
}

@article{bib50,
      author		= "Teixeira, J.R. and Tavares, L.A.M. and da, Silva, A.P.",
  title			= "Neurotoxicity and teratogenicity induced by carbendazim and ametryn in zebrafish: Implications for environmental and biological health",
  journal		= "Environ Toxicol Pharmacol",
  volume		= "122",
  pages			= "104947",
  year			= "2026",
}

@article{bib52,
      author		= "Huang, Y. and Li, Y. and Pan, H.",
  title			= "Global, regional, and national burden of neurological disorders in 204 countries and territories worldwide",
  journal		= "J Glob Health",
  volume		= "13",
  
  pages			= "04160",
  year			= "2023",
}

@article{bib53,
      author		= "Ding, C. and Wu, Y. and Chen, X.",
  title			= "Global, regional, and national burden and attributable risk factors of neurological disorders: The Global Burden of Disease study 1990-2019",
  journal		= "Front Public Health",
  volume		= "10",
  
  pages			= "952161",
  year			= "2019",
}

@article{bib54,
      author		= "Brown, D.L. and Staup, M. and Swanson, C.",
  title			= "Stereology of the Peripheral Nervous System",
  journal		= "Toxicol Pathol",
  volume		= "48",
  number="1",
  pages			= "37-38",
  year			= "2019",
}

@article{bib55,
      author		= "Chakif, D. and Furrer, J.",
  title			= "The impact of nutritional, environmental, and lifestyle factors on neurological disorders: therapeutic implications and mechanistic insights",
  journal		= "Front Pharmacol",
  volume		= "17",
  
  pages			= "1765786",
  year			= "2026",
}

@article{bib56,
      author		= "Lai, C.H. and Chou, C.Y. and Ch'ang, L.Y.",
  title			= "Identification of novel human genes evolutionarily conserved in Caenorhabditis elegans by comparative proteomics",
  journal		= "Genome Res",
  volume		= "10",
  pages			= "703-10",
  year			= "2000",
}

@article{bib57,
      author		= "Kaletta, T. and Hengartner, M.O.",
  title			= "Identification of novel human genes evolutionarily conserved in Caenorhabditis elegans by comparative proteomics",
  journal		= "Genome Res",
  volume		= "10",
  pages			= "703-10",
  year			= "2000",
}

@article{bib58,
      author		= "Ruszkiewicz, J. and Endig, L. and Güver, E.",
  title			= "Life-Cycle-Dependent Toxicities of Mono- and Bifunctional Alkylating Agents in the 3R-Compliant Model Organism C. elegans",
  journal		= "Cells",
  volume		= "12",
  number="23",
  pages			= "2728",
  year			= "2023",
}

@article{bib59,
      author		= "Pan, P. and Zhang, P. and Premachandran, S.",
  title			= "High-Resolution Imaging and Morphological Phenotyping of C. elegans through Stable Robotic Sample Rotation and Artificial Intelligence-Based 3-Dimensional Reconstruction",
  journal		= "Research",
  volume		= "7",
  
  pages			= "0513",
  year			= "2024",
}

@incollection{hart2006behavior,
    author    = "Hart, Anne C.",
    title     = "Behavior",
    booktitle = "WormBook: The Online Review of C. elegans Biology",
    publisher = "WormBook",
    address   = "Pasadena, CA",
    year      = "2006",
    doi       = "10.1895/wormbook.1.87.1",
}

@article{bib62,
      author		= "Stillman, Q.H. and Joseph, J.A. and Ahmed, J.",
  title			= "Protein mimetic 2D FAST rescues alpha synuclein aggregation mediated early and post disease Parkinson's phenotypes.",
  journal		= "Nat Commun",
  volume		= "15",
  number = "1",
  pages			= "3658",
  year			= "2024",
}

@article{bib63,
      author		= "Lynch, W.B. and Tschumi, C.W. and Sharpe, A.L.",
  title			= "Progressively disrupted somatodendritic morphology in dopamine neurons in a mouse Parkinson's model.",
  journal		= "Mov Disord",
  volume		= "33",
  number = "12",
  pages			= "1928-1937",
  year			= "2018",
}

@article{bib64,
      author		= "Ledonne, A. and Massaro, C.M. and Paldino, E.",
  title			= "Morpho-Functional Changes of Nigral Dopamine Neurons in an alpha-Synuclein Model of Parkinson's Disease",
  journal		= "Mov Disord",
  volume		= "38",
  number = "2",
  pages			= "256-266",
  year			= "2023",
}

@article{bib65,
      author		= "Miller, D.B.",
  title			= "Neurotoxicity of the pesticidal carbamates.",
  journal		= "Neurobehav Toxicol Teratol",
  volume		= "4",
  number = "6",
  pages			= "779-87",
  year			= "1982",
}

@article{bib66,
      author		= "Machemer, L.H. and Pickel, M.",
  title			= "Carbamate insecticides.",
  journal		= "Toxicology",
  volume		= "91",
  number = "1",
  pages			= "29-36",
  year			= "1994",
}

@article{bib67,
      author		= "Arellano, M. and Barnhill, L.M. and Kim, A.M.",
  title			= "Identification of pesticides associated with an increased risk of Parkinson's disease using a multi-screen approach.",
  journal		= "Environ Int",
  volume		= "208",
  pages			= "110087",
  year			= "2026",
}

@inproceedings{dosovitskiy2020image,
    title={An Image is Worth 16x16 Words: Transformers for Image Recognition at Scale},
    author={Alexey Dosovitskiy and Lucas Beyer and Alexander Kolesnikov and Dirk Weissenborn and Xiaohua Zhai and Thomas Unterthiner and Mostafa Dehghani and Matthias Minderer and Georg Heigold and Sylvain Gelly and Jakob Uszkoreit and Neil Houlsby},
    booktitle={International Conference on Learning Representations},
    year={2021},
}

@inproceedings{li2022exploring,
  title={Exploring plain vision transformer backbones for object detection},
  author={Li, Yanghao and Mao, Hanzi and Girshick, Ross and He, Kaiming},
  booktitle={European conference on computer vision},
  pages={280--296},
  year={2022},
}

@inproceedings{beyer2023flexivit,
  title={Flexivit: One model for all patch sizes},
  author={Beyer, Lucas and Izmailov, Pavel and Kolesnikov, Alexander and Caron, Mathilde and Kornblith, Simon and Zhai, Xiaohua and Minderer, Matthias and Tschannen, Michael and Alabdulmohsin, Ibrahim and Pavetic, Filip},
  booktitle={Proceedings of the IEEE/CVF Conference on Computer Vision and Pattern Recognition},
  pages={14496--14506},
  year={2023}
}

@inproceedings{he2022masked,
  title={Masked autoencoders are scalable vision learners},
  author={He, Kaiming and Chen, Xinlei and Xie, Saining and Li, Yanghao and Doll{\'a}r, Piotr and Girshick, Ross},
  booktitle={Proceedings of the IEEE/CVF conference on computer vision and pattern recognition},
  pages={16000--16009},
  year={2022}
}

@article{oquab2023dinov2,
  title={Dinov2: Learning robust visual features without supervision},
  author={Oquab, Maxime and Darcet, Timoth{\'e}e and Moutakanni, Th{\'e}o and Vo, Huy and Szafraniec, Marc and Khalidov, Vasil and Fernandez, Pierre and Haziza, Daniel and Massa, Francisco and El-Nouby, Alaaeldin and others},
  journal={arXiv preprint arXiv:2304.07193},
  year={2023}
}

@inproceedings{xie2021simmim,
  title={SimMIM: A Simple Framework for Masked Image Modeling},
  author={Xie, Zhenda and Zhang, Zheng and Cao, Yue and Lin, Yutong and Bao, Jianmin and Yao, Zhuliang and Dai, Qi and Hu, Han},
  booktitle={International Conference on Computer Vision and Pattern Recognition (CVPR)},
  year={2022}
}

@article{he2024foundation,
  title={Foundation model for advancing healthcare: challenges, opportunities and future directions},
  author={He, Yuting and Huang, Fuxiang and Jiang, Xinrui and Nie, Yuxiang and Wang, Minghao and Wang, Jiguang and Chen, Hao},
  journal={IEEE Reviews in Biomedical Engineering},
  volume={18},
  pages={172--191},
  year={2024}
}

@article{ma2024segment,
  title={Segment Anything in Medical Images},
  author={Ma, Jun and He, Yuting and Li, Feifei and Han, Lin and You, Chenyu and Wang, Bo},
  journal={Nature Communications},
  volume={15},
  pages={654},
  year={2024}
}

@article{zhang2024biomedclip,
  title={A multimodal biomedical foundation model trained from fifteen million image--text pairs},
  author={Zhang, Sheng and Xu, Yanbo and Usuyama, Naoto and Xu, Hanwen and Bagga, Jaspreet and Tinn, Robert and Preston, Sam and Rao, Rajesh and Wei, Mu and Valluri, Naveen and others},
  journal={Nejm Ai},
  volume={2},
  number={1},
  pages={AIoa2400640},
  year={2025}
}

@article{weigert2018content,
  title={Content-aware image restoration: pushing the limits of fluorescence microscopy},
  author={Weigert, Martin and Schmidt, Uwe and Boothe, Tobias and M{\"u}ller, Andreas and Dibrov, Alexandr and Jain, Akanksha and Wilhelm, Benjamin and Schmidt, Deborah and Broaddus, Coleman and Culley, Si{\^a}n and others},
  journal={Nature methods},
  volume={15},
  number={12},
  pages={1090--1097},
  year={2018}
}

@inproceedings{dutta2024recent,
  title={Recent Advancements in Microscopy Image Enhancement using Deep Learning: A Survey},
  author={Dutta, Debasish and Sonowal, Neeharika and Barauh, Risheraj and Chetia, Deepjyoti and Kalita, Sanjib Kr},
  booktitle={2024 IEEE International Conference on Computer Vision and Machine Intelligence (CVMI)},
  pages={1--7},
  year={2024}
}

@article{hollmann2023tabpfn,
  title={Accurate predictions on small data with a tabular foundation model},
  author={Hollmann, Noah and M{\"u}ller, Samuel and Purucker, Lennart and Krishnakumar, Arjun and K{\"o}rfer, Max and Hoo, Shi Bin and Schirrmeister, Robin Tibor and Hutter, Frank},
  journal={Nature},
  volume={637},
  number={8045},
  pages={319--326},
  year={2025}
}

@article{akyon2022sahi,
  title={Slicing Aided Hyper Inference and Fine-tuning for Small Object Detection},
  author={Akyon, Fatih Cagatay and Altinuc, Sinan Onur and Temizel, Alptekin},
  journal={2022 IEEE International Conference on Image Processing (ICIP)},
  doi={10.1109/ICIP46576.2022.9897990},
  pages={966-970},
  year={2022}
}

@misc{mmseg2020,
    title={{MMSegmentation}: OpenMMLab Semantic Segmentation Toolbox and Benchmark},
    author={MMSegmentation Contributors},
    howpublished = {\url{https://github.com/open-mmlab/mmsegmentation}},
    year={2020}
}

@article{mmdetection,
  title   = {{MMDetection}: Open MMLab Detection Toolbox and Benchmark},
  author  = {Chen, Kai and Wang, Jiaqi and Pang, Jiangmiao and Cao, Yuhang and
             Xiong, Yu and Li, Xiaoxiao and Sun, Shuyang and Feng, Wansen and
             Liu, Ziwei and Xu, Jiarui and Zhang, Zheng and Cheng, Dazhi and
             Zhu, Chenchen and Cheng, Tianheng and Zhao, Qijie and Li, Buyu and
             Lu, Xin and Zhu, Rui and Wu, Yue and Dai, Jifeng and Wang, Jingdong
             and Shi, Jianping and Ouyang, Wanli and Loy, Chen Change and Lin, Dahua},
  journal= {arXiv preprint arXiv:1906.07155},
  year={2019}
}

@inproceedings{chen2021mocov3,
  title={An empirical study of training self-supervised vision transformers},
  author={Chen, Xinlei and Xie, Saining and He, Kaiming},
  booktitle={Proceedings of the IEEE/CVF international conference on computer vision},
  pages={9640--9649},
  year={2021}
}

@inproceedings{assran2022msn,
  title={Masked siamese networks for label-efficient learning},
  author={Assran, Mahmoud and Caron, Mathilde and Misra, Ishan and Bojanowski, Piotr and Bordes, Florian and Vincent, Pascal and Joulin, Armand and Rabbat, Mike and Ballas, Nicolas},
  booktitle={European conference on computer vision},
  pages={456--473},
  year={2022}
}

@inproceedings{liu2023mixmae,
  title={MixMAE: Mixed and masked autoencoder for efficient pretraining of hierarchical vision transformers},
  author={Liu, Jihao and Huang, Xin and Zheng, Jinliang and Liu, Yu and Li, Hongsheng},
  booktitle={Proceedings of the IEEE/CVF Conference on Computer Vision and Pattern Recognition},
  pages={6252--6261},
  year={2023}
}

@inproceedings{zhang2023hivit,
  title={HiViT: A Simpler and More Efficient Design of Hierarchical Vision Transformer},
  author={Zhang, Xiaosong and Tian, Yunjie and Xie, Lingxi and Huang, Wei and Dai, Qi and Ye, Qixiang and Tian, Qi},
  booktitle={International Conference on Learning Representations},
  year={2023},
}

@article{oquab2024dinov2,
  title={DINOv2: Learning Robust Visual Features without Supervision},
  author={Oquab, Maxime and Darcet, Timoth{\'e}e and Moutakanni, Th{\'e}o and Vo, Huy and Szafraniec, Marc and Khalidov, Vasil and Fernandez, Pierre and Haziza, Daniel and Massa, Francisco and El-Nouby, Alaaeldin and others},
  journal={Transactions on Machine Learning Research Journal},
  year={2024}
}

@article{ma2024medsam,
  title={Segment anything in medical images},
  author={Ma, Jun and He, Yuting and Li, Feifei and Han, Lin and You, Chenyu and Wang, Bo},
  journal={Nature communications},
  volume={15},
  number={1},
  pages={654},
  year={2024}
}

@article{codella2024medimageinsight,
  title={Medimageinsight: An open-source embedding model for general domain medical imaging},
  author={Codella, Noel CF and Jin, Ying and Jain, Shrey and Gu, Yu and Lee, Ho Hin and Abacha, Asma Ben and Santamaria-Pang, Alberto and Guyman, Will and Sangani, Naiteek and Zhang, Sheng and others},
  journal={arXiv preprint arXiv:2410.06542},
  year={2024}
}

@article{fang2024eva,
  title={Eva-02: A visual representation for neon genesis},
  author={Fang, Yuxin and Sun, Quan and Wang, Xinggang and Huang, Tiejun and Wang, Xinlong and Cao, Yue},
  journal={Image and Vision Computing},
  volume={149},
  pages={105171},
  year={2024}
}

@article{tschannen2025siglip,
  title={Siglip 2: Multilingual vision-language encoders with improved semantic understanding, localization, and dense features},
  author={Tschannen, Michael and Gritsenko, Alexey and Wang, Xiao and Naeem, Muhammad Ferjad and Alabdulmohsin, Ibrahim and Parthasarathy, Nikhil and Evans, Talfan and Beyer, Lucas and Xia, Ye and Mustafa, Basil and others},
  journal={arXiv preprint arXiv:2502.14786},
  year={2025}
}

\clearpage

\begin{appendices}

\renewcommand{\theHfigure}{S\arabic{figure}}

\section{Extended Data}\label{secA1}

\begin{figure*}[htbp]
    \centering
    \includegraphics[width=1.0\textwidth, trim=0cm 14cm 0cm 0cm, clip]{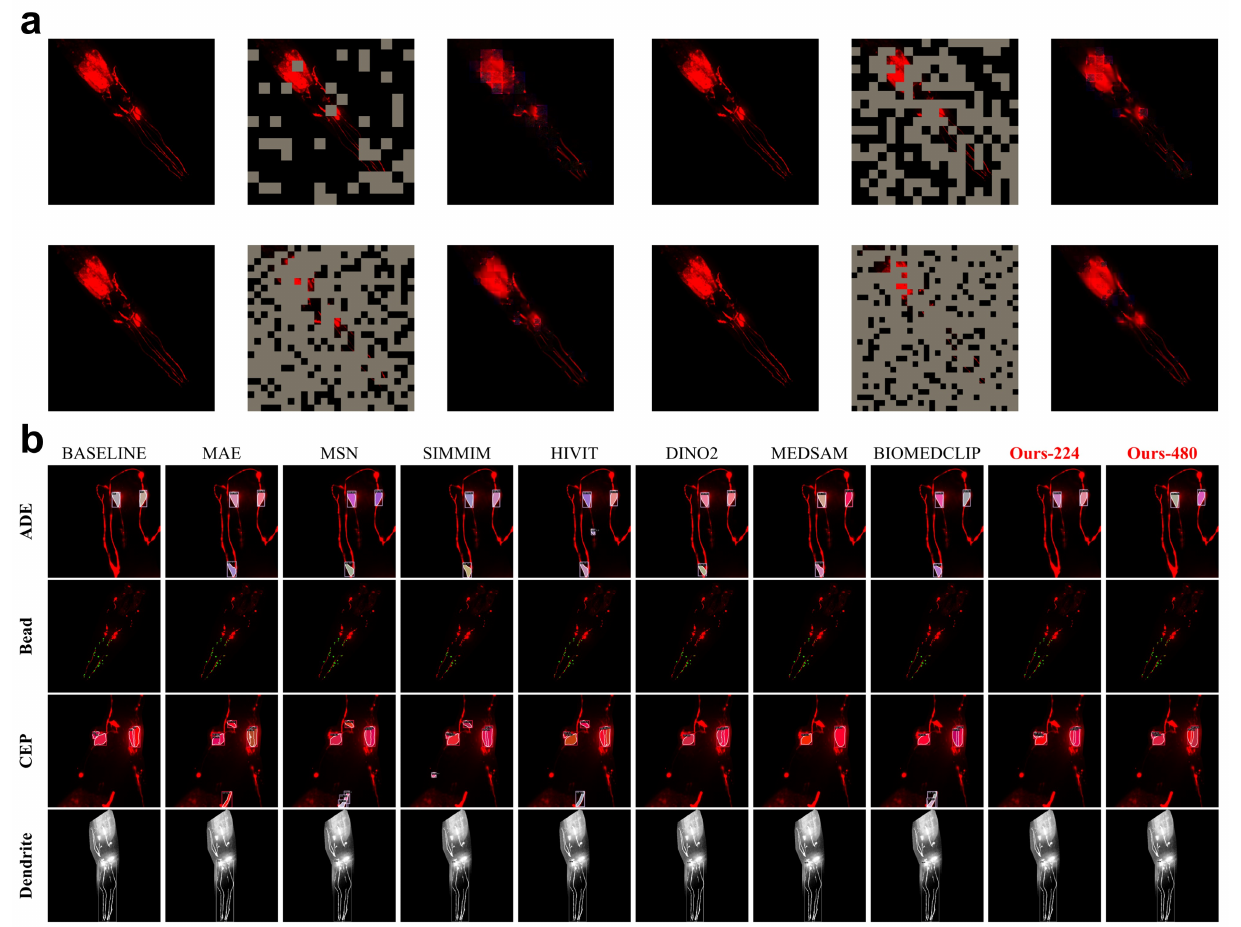}
    \caption{\textbf{Qualitative results of our proposed framework.} \textbf{a}, Reconstructed confocal microscopy images of \textit{C. elegans} from highly masked inputs at multiple scales in the self-supervised pretext task. Despite the limited visibility of patches, the model successfully infers and rebuilds continuous neural anatomical structures and fine-grained morphological details. \textbf{b}, Qualitative comparison of downstream task performance against baseline and state-of-the-art foundation models. The visualizations highlight the model predictions for specific neurons (ADE, CEP), dendritic structures, and beading/punctate morphology (a critical marker for degenerative changes, rather than subcellular artifacts). Compared to other methods, our models (Ours-224 and Ours-480) demonstrate superior accuracy and boundary precision in localizing and segmenting these intricate biological features.}
    \label{fig_A1}
\end{figure*}

\begin{figure*}[htbp]
    \centering
    \includegraphics[width=1.0\textwidth, trim=0cm 15cm 0cm 0cm, clip]{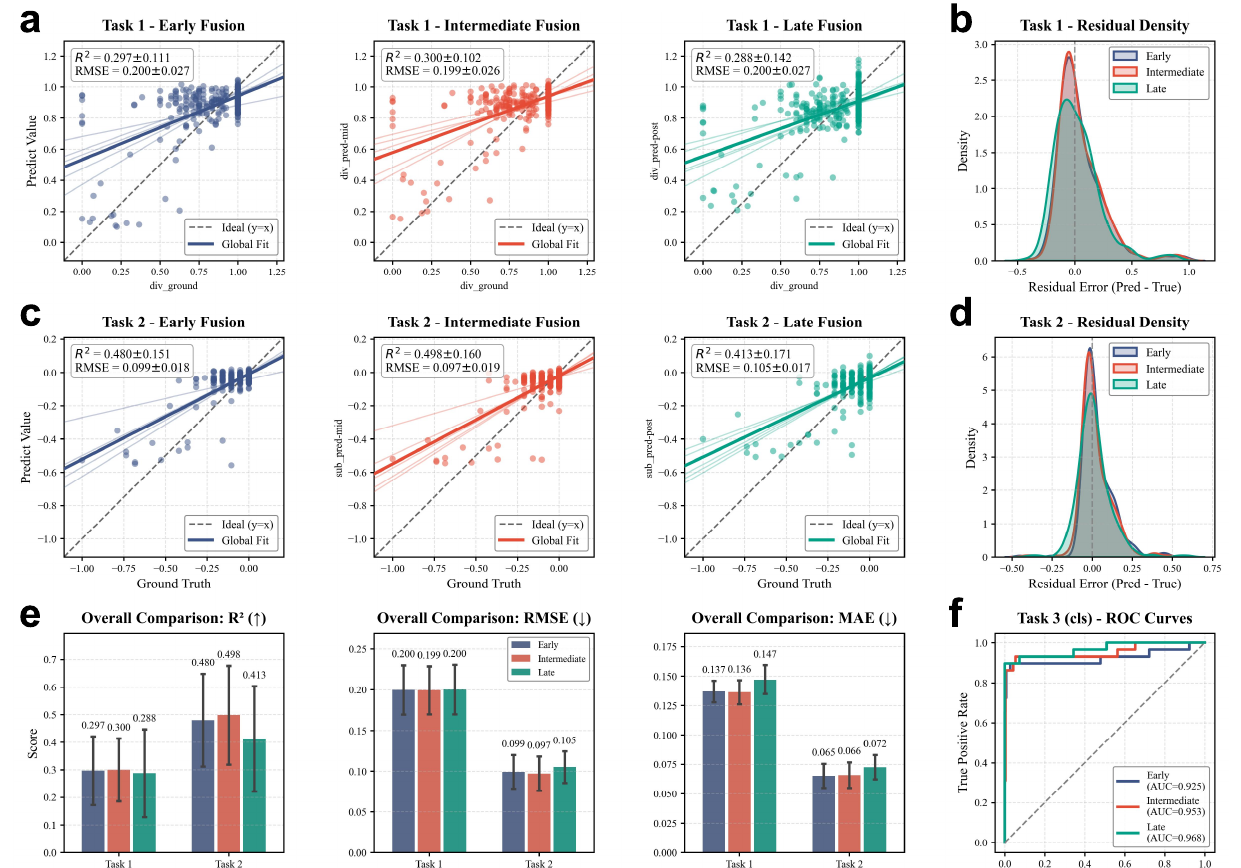}
    \caption{\textbf{Predictive performance and evaluation metrics for regression tasks under different fusion strategies.} 
    (a--b) Display the scatter plots of predicted values versus ground truth (including fold-level and global linear fits) and the corresponding residual error density distributions for Task 1, respectively. 
    (c--d) Present the equivalent scatter plots and residual error density distributions for Task 2. 
    (e) Summarizes the overall quantitative regression metrics ($R^2$, RMSE, and MAE) across all methods, with error bars indicating the standard deviation across the 5 folds. 
    (f) Illustrates the Receiver Operating Characteristic (ROC) curves and corresponding Area Under the Curve (AUC) scores for the Task 3 classification under different fusion strategies, calculated by pooling the predictions from all 5 folds.}
    \label{fig_A2}
\end{figure*}

\begin{sidewaystable}[htbp]
    \centering
    \caption{Ablation study of the proposed scale-adaptive framework. We progressively evaluate the impact of Scale-adaptive (SA) MIM Pre-training, Length-Constrained (LC) Masking, and Dual-scale Fusion on downstream tasks. Performance is evaluated using Accuracy for classification, Dice score for semantic segmentation, bounding box mAP (mAP$^{\text{box}}$) for object detection, and mask mAP (mAP$^{\text{mask}}$) for instance segmentation.}
    \label{tab:table_s1}
    \footnotesize
    \setlength{\tabcolsep}{4pt}
    \begin{tabular}{lccc ccc c c cc}
    \toprule
    \multirow{2}{*}{Model Variant} & \multirow{2}{*}{SA MIM} & \multirow{2}{*}{LC Mask} & \multirow{2}{*}{Dual-scale} & \multicolumn{3}{c}{Classification} & Sem. Seg. & Obj. Det. & \multicolumn{2}{c}{Inst. Seg.} \\
    \cmidrule(lr){5-7} \cmidrule(lr){8-8} \cmidrule(lr){9-9} \cmidrule(lr){10-11}
     & & & & Break & Arb. & Bend & Bead & Dendrite & ADE & CEP \\
    \midrule
    Baseline (ViT-B MAE) & $\times$ & $\times$ & $\times$ & 86.97$_{\pm1.03}$ & 81.04$_{\pm1.46}$ & 75.26$_{\pm3.00}$ & 75.60$_{\pm0.41}$ & 87.40$_{\pm0.86}$ & 59.32$_{\pm0.40}$ & 66.40$_{\pm1.02}$ \\
    $+$ SA MIM             & $\checkmark$ & $\times$ & $\times$ & 87.73$_{\pm0.51}$ & 81.85$_{\pm2.53}$ & 79.21$_{\pm1.10}$ & 76.51$_{\pm0.55}$ & 89.16$_{\pm0.63}$ & 60.18$_{\pm0.79}$ & 67.08$_{\pm0.29}$ \\
    $+$ LC Mask            & $\checkmark$ & $\checkmark$ & $\times$ & 87.77$_{\pm0.49}$ & 82.94$_{\pm1.15}$ & 80.79$_{\pm1.50}$ & 77.43$_{\pm0.47}$ & 89.26$_{\pm1.27}$ & 61.32$_{\pm1.12}$ & 69.60$_{\pm0.63}$ \\
    $+$ Dual-scale (Full)  & $\checkmark$ & $\checkmark$ & $\checkmark$ & \textbf{87.96}$_{\pm0.46}$ & \textbf{83.79}$_{\pm0.76}$ & \textbf{81.05}$_{\pm2.73}$ & \textbf{78.22}$_{\pm0.38}$ & \textbf{89.84}$_{\pm0.33}$ & \textbf{62.70}$_{\pm1.69}$ & \textbf{69.70}$_{\pm0.78}$ \\
    \bottomrule
    \end{tabular}
\end{sidewaystable}

\begin{figure}[htbp]
    \centering
    \includegraphics[width=1.0\textwidth, trim=0cm 2.5cm 0cm 2cm, clip]{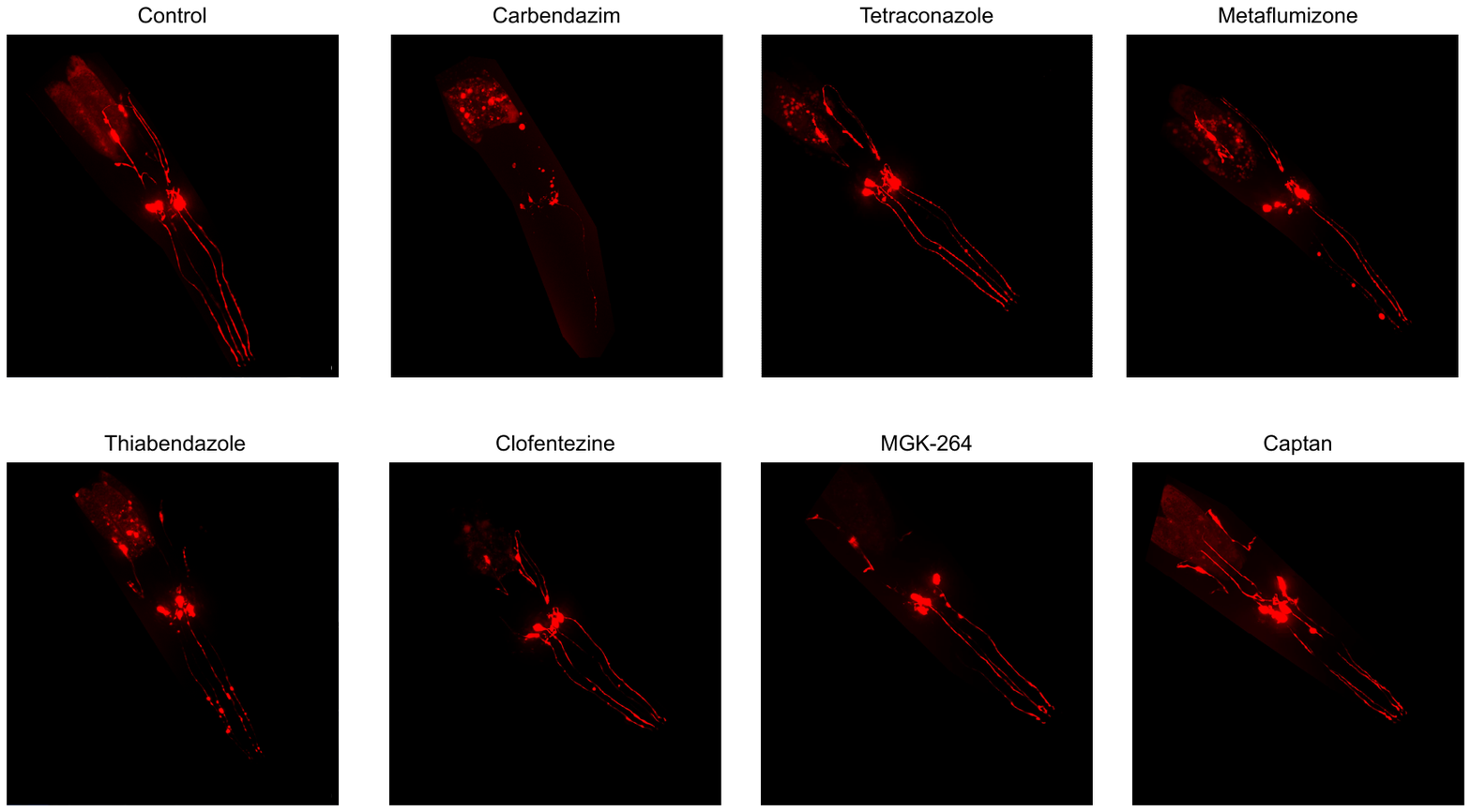}
    \caption{Representative head neuron fluorescence images of \textit{C. elegans} exposed to five top-ranked pesticides excluding carbendazim and thiabendazole.}
    \label{fig_A3}
\end{figure}

\begin{figure}[htbp]
    \centering
    \includegraphics[width=1.0\linewidth, trim=0cm 2.2cm 1.5cm 2cm, clip]{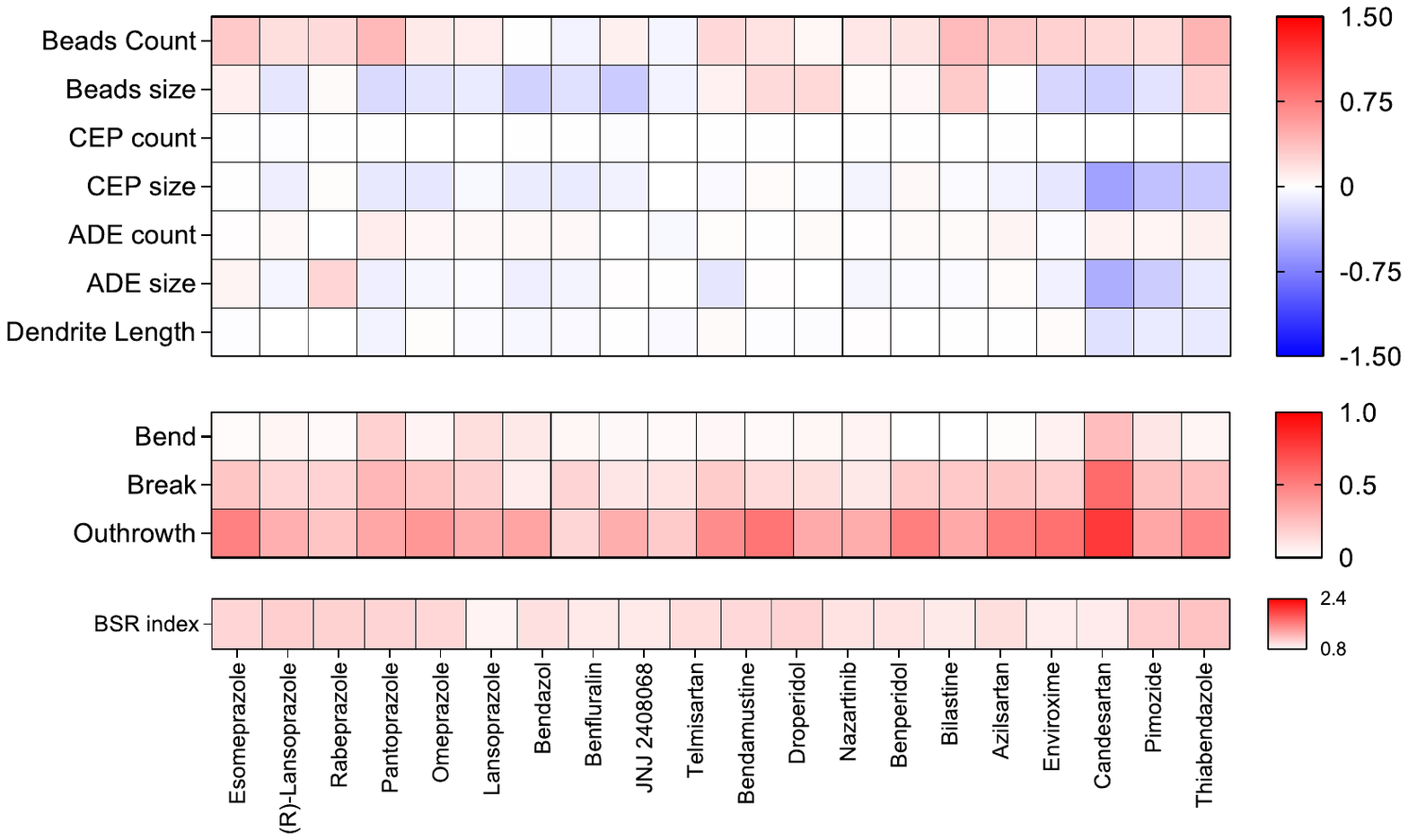}
    \caption{Dopaminergic neuron injury patterns and BSR index distribution following benzimidazole exposure in \textit{C. elegans}.}
    \label{fig_A4}
\end{figure}



\end{appendices}


\end{document}